\documentclass[twocolumn]{aastex701}

\usepackage{hyperref}
\usepackage{verbatim}
\usepackage{xcolor}

\usepackage{multirow}
\usepackage{amsmath}
\usepackage{float}
\usepackage{adjustbox}
\usepackage{footnote}

\usepackage{xspace}
\newcommand{\rp}{$r_\parallel$\xspace}
\newcommand{\rt}{$r_\perp$\xspace}
\newcommand{\mpc}{~$h^{-1}$~Mpc\xspace}
\newcommand{\civ}{C\,{\sc iv}\xspace}
\newcommand{\siiv}{Si\,{\sc iv}\xspace}

\defcitealias{desi-dr2-bao}{\texttt{DESI-2025-I}}
\newcommand{\lyabao}{\citetalias{desi-dr2-bao}\xspace}

\defcitealias{desi-dr2-gal}{\texttt{DESI-2025-II}}
\newcommand{\galbao}{\citetalias{desi-dr2-gal}\xspace}

\begin{document}

\shortauthors{Bault et al.}
\shorttitle{DESI DR2 CIV BAO}

\title{Baryon Acoustic Oscillations from the \civ Forest with DESI DR2}



\author[0000-0002-9964-1005, gname='Abby', sname='Bault']{Abby~Bault}
\affiliation{Lawrence Berkeley National Laboratory, 1 Cyclotron Road, Berkeley, CA 94720, USA}
\email{abault@lbl.gov}

\author[0000-0002-2169-0595, gname='Andrei', sname='Cuceu']{Andrei~Cuceu}
\affiliation{Lawrence Berkeley National Laboratory, 1 Cyclotron Road, Berkeley, CA 94720, USA}
\affiliation{NASA Einstein Fellow}
\email{acuceu@lbl.gov}

\author[0000-0001-9822-6793, gname='Julien', sname='Guy']{Julien~Guy}
\affiliation{Lawrence Berkeley National Laboratory, 1 Cyclotron Road, Berkeley, CA 94720, USA}
\email{jguy@lbl.gov}

\author[gname='Jessica Nicole', sname='Aguilar']{J.~Aguilar}
\affiliation{Lawrence Berkeley National Laboratory, 1 Cyclotron Road, Berkeley, CA 94720, USA}
\email{jaguilar@lbl.gov}

\author[0000-0001-6098-7247, gname='Steven', sname='Ahlen']{S.~Ahlen}
\affiliation{Department of Physics, Boston University, 590 Commonwealth Avenue, Boston, MA 02215 USA}
\email{ahlen@bu.edu}

\author[0000-0001-9712-0006, gname='Davide', sname='Bianchi']{D.~Bianchi}
\affiliation{Dipartimento di Fisica ``Aldo Pontremoli'', Universit\`a degli Studi di Milano, Via Celoria 16, I-20133 Milano, Italy}
\affiliation{INAF-Osservatorio Astronomico di Brera, Via Brera 28, 20122 Milano, Italy}
\email{davide.bianchi1@unimi.it}

\author[0000-0002-8934-0954, gname='Allyson', sname='Brodzeller']{A.~Brodzeller}
\affiliation{Lawrence Berkeley National Laboratory, 1 Cyclotron Road, Berkeley, CA 94720, USA}
\email{AllysonBrodzeller@lbl.gov}

\author[gname='David', sname='Brooks']{D.~Brooks}
\affiliation{Department of Physics \& Astronomy, University College London, Gower Street, London, WC1E 6BT, UK}
\email{david.brooks@ucl.ac.uk}

\author[gname='Rebecca', sname='Canning']{R.~Canning}
\affiliation{Institute of Cosmology and Gravitation, University of Portsmouth, Dennis Sciama Building, Portsmouth, PO1 3FX, UK}
\email{becky.canning@port.ac.uk}

\author[0000-0001-8996-4874, gname='Edmond', sname='Chaussidon']{E.~Chaussidon}
\affiliation{Lawrence Berkeley National Laboratory, 1 Cyclotron Road, Berkeley, CA 94720, USA}
\email{echaussidon@lbl.gov}

\author[gname='Todd', sname='Claybaugh']{T.~Claybaugh}
\affiliation{Lawrence Berkeley National Laboratory, 1 Cyclotron Road, Berkeley, CA 94720, USA}
\email{tmclaybaugh@lbl.gov}

\author[0000-0003-3660-4028, gname='Roger', sname='de Belsunce']{R.~de Belsunce}
\affiliation{Lawrence Berkeley National Laboratory, 1 Cyclotron Road, Berkeley, CA 94720, USA}
\email{belsunce@mit.edu}

\author[0000-0002-1769-1640, gname='Axel ', sname='de la Macorra']{A.~de la Macorra}
\affiliation{Instituto de F\'{\i}sica, Universidad Nacional Aut\'{o}noma de M\'{e}xico,  Circuito de la Investigaci\'{o}n Cient\'{\i}fica, Ciudad Universitaria, Cd. de M\'{e}xico  C.~P.~04510,  M\'{e}xico}
\email{macorra@fisica.unam.mx}

\author[0000-0002-4928-4003, gname='Arjun', sname='Dey']{Arjun~Dey}
\affiliation{NSF NOIRLab, 950 N. Cherry Ave., Tucson, AZ 85719, USA}
\email{arjun.dey@noirlab.edu}

\author[gname='Peter', sname='Doel']{P.~Doel}
\affiliation{Department of Physics \& Astronomy, University College London, Gower Street, London, WC1E 6BT, UK}
\email{apd@star.ucl.ac.uk}

\author[0000-0003-4992-7854, gname='Simone', sname='Ferraro']{S.~Ferraro}
\affiliation{Lawrence Berkeley National Laboratory, 1 Cyclotron Road, Berkeley, CA 94720, USA}
\affiliation{University of California, Berkeley, 110 Sproul Hall \#5800 Berkeley, CA 94720, USA}
\email{sferraro@lbl.gov}

\author[0000-0002-3033-7312, gname='Andreu', sname='Font-Ribera']{A.~Font-Ribera}
\affiliation{Instituci\'{o} Catalana de Recerca i Estudis Avan\c{c}ats, Passeig de Llu\'{\i}s Companys, 23, 08010 Barcelona, Spain}
\affiliation{Institut de F\'{i}sica d’Altes Energies (IFAE), The Barcelona Institute of Science and Technology, Edifici Cn, Campus UAB, 08193, Bellaterra (Barcelona), Spain}
\email{afont@ifae.es}

\author[0000-0002-2890-3725, gname='Jaime E.', sname='Forero-Romero']{J.~E.~Forero-Romero}
\affiliation{Departamento de F\'isica, Universidad de los Andes, Cra. 1 No. 18A-10, Edificio Ip, CP 111711, Bogot\'a, Colombia}
\affiliation{Observatorio Astron\'omico, Universidad de los Andes, Cra. 1 No. 18A-10, Edificio H, CP 111711 Bogot\'a, Colombia}
\email{je.forero@uniandes.edu.co}

\author[0000-0001-9632-0815, gname='Enrique', sname='Gaztañaga']{E.~Gaztañaga}
\affiliation{Institut d'Estudis Espacials de Catalunya (IEEC), c/ Esteve Terradas 1, Edifici RDIT, Campus PMT-UPC, 08860 Castelldefels, Spain}
\affiliation{Institute of Cosmology and Gravitation, University of Portsmouth, Dennis Sciama Building, Portsmouth, PO1 3FX, UK}
\affiliation{Institute of Space Sciences, ICE-CSIC, Campus UAB, Carrer de Can Magrans s/n, 08913 Bellaterra, Barcelona, Spain}
\email{gaztanaga@gmail.com}

\author[0000-0003-3142-233X, gname='Satya ', sname='Gontcho A Gontcho']{S.~Gontcho A Gontcho}
\affiliation{Lawrence Berkeley National Laboratory, 1 Cyclotron Road, Berkeley, CA 94720, USA}
\affiliation{University of Virginia, Department of Astronomy, Charlottesville, VA 22904, USA}
\email{satya@virginia.edu}

\author[0000-0003-2561-5733, gname='Calum', sname='Gordon']{C.~Gordon}
\affiliation{Department of Physics \& Astronomy, University College London, Gower Street, London, WC1E 6BT, UK}
\email{cgordon@ifae.es}

\author[0000-0002-0676-3661, gname='Dylan', sname='Green']{D.~Green}
\affiliation{Lawrence Berkeley National Laboratory, 1 Cyclotron Road, Berkeley, CA 94720, USA}
\email{dylangreen@lbl.gov}

\author[gname='Gaston', sname='Gutierrez']{G.~Gutierrez}
\affiliation{Fermi National Accelerator Laboratory, PO Box 500, Batavia, IL 60510, USA}
\email{gaston@fnal.gov}

\author[0000-0003-1197-0902, gname='ChangHoon', sname='Hahn']{C.~Hahn}
\affiliation{Department of Astronomy, The University of Texas at Austin, 2515 Speedway Boulevard, Austin, TX 78712, USA}
\email{changhoon.hahn@utexas.edu}

\author[0000-0002-9136-9609, gname='Hiram K.', sname='Herrera-Alcantar']{H.~K.~Herrera-Alcantar}
\affiliation{Institut d'Astrophysique de Paris. 98 bis boulevard Arago. 75014 Paris, France}
\affiliation{IRFU, CEA, Universit\'{e} Paris-Saclay, F-91191 Gif-sur-Yvette, France}
\email{herreraa@iap.fr}

\author[0000-0002-6550-2023, gname='Klaus', sname='Honscheid']{K.~Honscheid}
\affiliation{Center for Cosmology and AstroParticle Physics, The Ohio State University, 191 West Woodruff Avenue, Columbus, OH 43210, USA}
\affiliation{Department of Physics, The Ohio State University, 191 West Woodruff Avenue, Columbus, OH 43210, USA}
\affiliation{The Ohio State University, Columbus, 43210 OH, USA}
\email{kh@physics.osu.edu}

\author[0000-0002-6024-466X, gname='Mustapha', sname='Ishak']{M.~Ishak}
\affiliation{Department of Physics, The University of Texas at Dallas, 800 W. Campbell Rd., Richardson, TX 75080, USA}
\email{mishak@utdallas.edu}

\author[0000-0003-0201-5241, gname='Dick', sname='Joyce']{R.~Joyce}
\affiliation{NSF NOIRLab, 950 N. Cherry Ave., Tucson, AZ 85719, USA}
\email{richard.joyce@noirlab.edu}

\author[0000-0002-0000-2394, gname='Stephanie', sname='Juneau']{S.~Juneau}
\affiliation{NSF NOIRLab, 950 N. Cherry Ave., Tucson, AZ 85719, USA}
\email{stephanie.juneau@noirlab.edu}

\author[0000-0002-8828-5463, gname='David', sname='Kirkby']{D.~Kirkby}
\affiliation{Department of Physics and Astronomy, University of California, Irvine, 92697, USA}
\email{dkirkby@uci.edu}

\author[0000-0001-6356-7424, gname='Anthony', sname='Kremin']{A.~Kremin}
\affiliation{Lawrence Berkeley National Laboratory, 1 Cyclotron Road, Berkeley, CA 94720, USA}
\email{akremin@lbl.gov}

\author[0000-0002-6731-9329, gname='Claire', sname='Lamman']{C.~Lamman}
\affiliation{The Ohio State University, Columbus, 43210 OH, USA}
\email{lamman.1@osu.edu}

\author[0000-0003-1838-8528, gname='Martin', sname='Landriau']{M.~Landriau}
\affiliation{Lawrence Berkeley National Laboratory, 1 Cyclotron Road, Berkeley, CA 94720, USA}
\email{mlandriau@lbl.gov}

\author[0000-0001-7178-8868, gname='Laurent', sname='Le Guillou']{L.~Le~Guillou}
\affiliation{Sorbonne Universit\'{e}, CNRS/IN2P3, Laboratoire de Physique Nucl\'{e}aire et de Hautes Energies (LPNHE), FR-75005 Paris, France}
\email{llg@lpnhe.in2p3.fr}

\author[0000-0003-1887-1018, gname='Michael', sname='Levi']{M.~E.~Levi}
\affiliation{Lawrence Berkeley National Laboratory, 1 Cyclotron Road, Berkeley, CA 94720, USA}
\email{melevi@lbl.gov}

\author[0000-0003-4962-8934, gname='Marc', sname='Manera']{M.~Manera}
\affiliation{Departament de F\'{i}sica, Serra H\'{u}nter, Universitat Aut\`{o}noma de Barcelona, 08193 Bellaterra (Barcelona), Spain}
\affiliation{Institut de F\'{i}sica d’Altes Energies (IFAE), The Barcelona Institute of Science and Technology, Edifici Cn, Campus UAB, 08193, Bellaterra (Barcelona), Spain}
\email{mmanera@ifae.es}

\author[0000-0002-4279-4182, gname='Paul', sname='Martini']{P.~Martini}
\affiliation{Center for Cosmology and AstroParticle Physics, The Ohio State University, 191 West Woodruff Avenue, Columbus, OH 43210, USA}
\affiliation{Department of Astronomy, The Ohio State University, 4055 McPherson Laboratory, 140 W 18th Avenue, Columbus, OH 43210, USA}
\affiliation{The Ohio State University, Columbus, 43210 OH, USA}
\email{martini.10@osu.edu}

\author[0000-0002-1125-7384, gname='Aaron', sname='Meisner']{A.~Meisner}
\affiliation{NSF NOIRLab, 950 N. Cherry Ave., Tucson, AZ 85719, USA}
\email{aaron.meisner@noirlab.edu}

\author[gname='Ramon', sname='Miquel']{R.~Miquel}
\affiliation{Instituci\'{o} Catalana de Recerca i Estudis Avan\c{c}ats, Passeig de Llu\'{\i}s Companys, 23, 08010 Barcelona, Spain}
\affiliation{Institut de F\'{i}sica d’Altes Energies (IFAE), The Barcelona Institute of Science and Technology, Edifici Cn, Campus UAB, 08193, Bellaterra (Barcelona), Spain}
\email{rmiquel@ifae.es}

\author[0000-0002-2733-4559, gname='John', sname='Moustakas']{J.~Moustakas}
\affiliation{Department of Physics and Astronomy, Siena University, 515 Loudon Road, Loudonville, NY 12211, USA}
\email{jmoustakas@siena.edu}

\author[gname='Andrea ', sname='Muñoz-Gutiérrez']{A.~Muñoz-Gutiérrez}
\affiliation{Instituto de F\'{\i}sica, Universidad Nacional Aut\'{o}noma de M\'{e}xico,  Circuito de la Investigaci\'{o}n Cient\'{\i}fica, Ciudad Universitaria, Cd. de M\'{e}xico  C.~P.~04510,  M\'{e}xico}
\email{andreamgtz@ciencias.unam.mx}

\author[0000-0001-9070-3102, gname='Seshadri', sname='Nadathur']{S.~Nadathur}
\affiliation{Institute of Cosmology and Gravitation, University of Portsmouth, Dennis Sciama Building, Portsmouth, PO1 3FX, UK}
\email{seshadri.nadathur@port.ac.uk}

\author[0000-0003-3188-784X, gname='Nathalie', sname='Palanque-Delabrouille']{N.~Palanque-Delabrouille}
\affiliation{IRFU, CEA, Universit\'{e} Paris-Saclay, F-91191 Gif-sur-Yvette, France}
\affiliation{Lawrence Berkeley National Laboratory, 1 Cyclotron Road, Berkeley, CA 94720, USA}
\email{npalanque-delabrouille@lbl.gov}

\author[0000-0002-0644-5727, gname='Will', sname='Percival']{W.~J.~Percival}
\affiliation{Department of Physics and Astronomy, University of Waterloo, 200 University Ave W, Waterloo, ON N2L 3G1, Canada}
\affiliation{Perimeter Institute for Theoretical Physics, 31 Caroline St. North, Waterloo, ON N2L 2Y5, Canada}
\affiliation{Waterloo Centre for Astrophysics, University of Waterloo, 200 University Ave W, Waterloo, ON N2L 3G1, Canada}
\email{will.percival@uwaterloo.ca}

\author[0000-0003-0247-8991, gname='Matthew', sname='Pieri']{Matthew~M.~Pieri}
\affiliation{Aix Marseille Univ, CNRS, CNES, LAM, Marseille, France}
\email{matthew.pieri@lam.fr}

\author[gname='Claire', sname='Poppett']{C.~Poppett}
\affiliation{Lawrence Berkeley National Laboratory, 1 Cyclotron Road, Berkeley, CA 94720, USA}
\affiliation{Space Sciences Laboratory, University of California, Berkeley, 7 Gauss Way, Berkeley, CA  94720, USA}
\affiliation{University of California, Berkeley, 110 Sproul Hall \#5800 Berkeley, CA 94720, USA}
\email{clpoppett@lbl.gov}

\author[0000-0001-7145-8674, gname='Francisco', sname='Prada']{F.~Prada}
\affiliation{Instituto de Astrof\'{i}sica de Andaluc\'{i}a (CSIC), Glorieta de la Astronom\'{i}a, s/n, E-18008 Granada, Spain}
\email{fprada@iaa.es}

\author[0000-0001-6979-0125, gname='Ignasi', sname='Pérez-Ràfols']{I.~P\'erez-R\`afols}
\affiliation{Departament de F\'isica, EEBE, Universitat Polit\`ecnica de Catalunya, c/Eduard Maristany 10, 08930 Barcelona, Spain}
\email{ignasi.perez.rafols@upc.edu}

\author[gname='Graziano', sname='Rossi']{G.~Rossi}
\affiliation{Department of Physics and Astronomy, Sejong University, 209 Neungdong-ro, Gwangjin-gu, Seoul 05006, Republic of Korea}
\email{graziano@sejong.ac.kr}

\author[0000-0002-9646-8198, gname='Eusebio', sname='Sanchez']{E.~Sanchez}
\affiliation{CIEMAT, Avenida Complutense 40, E-28040 Madrid, Spain}
\email{eusebio.sanchez@ciemat.es}

\author[gname='David', sname='Schlegel']{D.~Schlegel}
\affiliation{Lawrence Berkeley National Laboratory, 1 Cyclotron Road, Berkeley, CA 94720, USA}
\email{djschlegel@lbl.gov}

\author[0000-0002-6588-3508, gname='Hee-Jong', sname='Seo']{H.~Seo}
\affiliation{Department of Physics \& Astronomy, Ohio University, 139 University Terrace, Athens, OH 45701, USA}
\email{seoh@ohio.edu}

\author[0000-0002-3461-0320, gname='Joseph Harry', sname='Silber']{J.~Silber}
\affiliation{Lawrence Berkeley National Laboratory, 1 Cyclotron Road, Berkeley, CA 94720, USA}
\email{jhsilber@lbl.gov}

\author[gname='David', sname='Sprayberry']{D.~Sprayberry}
\affiliation{NSF NOIRLab, 950 N. Cherry Ave., Tucson, AZ 85719, USA}
\email{david.sprayberry@noirlab.edu}

\author[0000-0003-1704-0781, gname='Gregory', sname='Tarlé']{G.~Tarl\'{e}}
\affiliation{University of Michigan, 500 S. State Street, Ann Arbor, MI 48109, USA}
\email{gtarle@umich.edu}

\author[gname='Benjamin Alan', sname='Weaver']{B.~A.~Weaver}
\affiliation{NSF NOIRLab, 950 N. Cherry Ave., Tucson, AZ 85719, USA}
\email{benjamin.weaver@noirlab.edu}

\correspondingauthor{Abby Bault}
\email{abault@lbl.gov}


\begin{abstract}
We present a measurement of Baryon Acoustic Oscillations (BAO) in the cross-correlation of triply ionized carbon (\civ) absorption with the positions of quasars (QSO) and Emission Line Galaxies (ELG). We use quasars and ELGs from the second data release (DR2) of the Dark Energy Spectroscopic Instrument (DESI) survey. Our data sample consists of 2.5 million quasars, 3.1 million ELGs, and the \civ absorption is measured along the line of sight of 1.5 million high redshift quasars with $z > 1.3$. We measure the isotropic BAO signal at 4.2$\sigma$ for the CIV$\times$QSO cross-correlation. This translates into a 3.0\% precision measurement of the ratio of the isotropic distance scale, $D_{\rm V}$, and the sound horizon at the drag epoch, $r_{\rm d}$, with $D_{\rm V}/r_{\rm d}(z_{\rm eff} = 1.92) = 30.3 \pm 0.9$. We make the first detection of the BAO feature in the CIV$\times$ELG cross-correlation at a significance of 2.5$\sigma$ and find $D_{\rm V}/r_{\rm d}(z_{\rm eff} = 1.47) = 24.6 \pm 1.0$.
\end{abstract}

\section{Introduction}
\label{sec:intro}

Baryon Acoustic Oscillations (BAO) have been used to study the expansion history of the Universe for two decades \citep{weinberg2012,eisenstein2007}. These oscillations are sound waves that formed due to perturbations in the primordial plasma in the early universe. The sound waves were ``frozen" into the matter distribution during recombination and can be seen in the distribution of matter at large scales in the universe \citep{weinberg2012, hu1996,eisenstein1998,eisenstein2007,eisenstein2008}. The BAO scale, once normalized to the sound horizon $r_{\rm d}$, is a standard cosmological ruler and serves as a tool for probing the expansion of the universe.

Typical BAO measurements use galaxies as tracers of the matter distribution through the two-point correlation function. This correlation function quantifies the probability of two tracers being separated by a given distance, generally measured in comoving coordinates for a choice of fiducial cosmology. Using various tracers at different redshifts for measurements of BAO probe the Universe at different energy densities, allowing for comparisons between a more matter dominated Universe versus a dark energy dominated Universe.

The precision of BAO measurements with galaxies at high redshifts is restricted by the low number density of sources that are bright enough to secure their redshifts, and by the absence of bright emission lines redshifted to wavelengths shorter than 1 micron for instruments equipped with silicon Charge-Coupled Device (CCD) detectors. To mitigate this, spectroscopic surveys have incorporated other tracers, such as high redshift quasars (with $z \geq 2$) and the Lyman-$\alpha$ forest, that allow for precise BAO measurements at redshifts that were unattainable with galaxy surveys alone.

The most recent set of BAO measurements comes from the Dark Energy Spectroscopic Instrument (DESI; \citealt{desi-2016b,desi-inst2022}). With the Data Release 1 (DR1; \citealt{desi-dr1}), DESI released BAO measurements with galaxies and quasars \citep{desi-dr1-kp4} and the Lyman-$\alpha$ forest \citep{desi-dr1-kp6}, and cosmological constraints from full-shape measurements \citep{desi-dr1-kp7b} and from BAO \citep{desi-dr1-kp7a}. Most recently, DESI released updated analyses using Data Release 2 (DR2) for measurements of BAO with the Lyman-$\alpha$ forest \citep[][hereafter \lyabao]{desi-dr2-bao} and cosmological constraints with galaxies and quasars \citep[][hereafter \galbao]{desi-dr2-gal}. These updated results confirm what was found in the DESI Year 1 results that dark energy may be evolving.

Galaxies, such as Emission Line Galaxies (ELG) and Luminous Red Galaxies (LRG), and the Lyman-$\alpha$ forest are not the only tracers that can be used to measure BAO. In fact, as efforts to observe the universe at high redshift have increased, BAO measurements with other tracers, like Lyman-break galaxies (LBG) and Lyman-alpha emitting galaxies (LAE), have become increasingly common as instruments become more powerful, allowing for the precision and density needed for fainter tracers \citep{hiram2025, lae-desi,sailer2021,ferraro2022,ouchi2020}. 

One tracer that has been infrequently explored is the doublet transition of triply ionized carbon (\civ). Often classified as a metal or a contaminant in Lyman-$\alpha$ forest analyses, \civ offers a unique window for measurements at different redshifts than those with galaxies, quasars, and the Lyman-$\alpha$ forest. \civ has a strong doublet ($\lambda_{\rm RF}$ = 1548.20, 1550.78 \AA) and is the dominant absorber between the Lyman-$\alpha$ and \civ emission lines \citep{pieri2014a}. In optical spectra, the Lyman-$\alpha$ forest can only be seen at redshifts $z > 2$, however, DESI can observe the \civ doublet down to $z \sim 1.3$. This allows for measurements at various effective redshifts when correlated with different tracer positions.

When cross-correlated with quasars, a BAO measurement with the \civ forest allows for additional constraints on cosmological parameters at an effective redshift of $z_{\rm eff} \sim 2$, directly in the redshift gap between the measurement from the QSO auto-correlation and the Lyman-$\alpha$ forest (see Figure 6 of \galbao). Cross-correlating the \civ forest with the DESI Emission Line Galaxy (ELG) sample would allow for an additional measurement around $z_{\rm eff} \sim 1.5$. The DESI ELG sample probes the Universe over the $0.6 < z < 1.6$ range and brings the tightest cosmological constraints from DESI \citep{desi-ts-elg}.

This work is largely based on \cite{blomqvist2018}, who presented the first BAO measurement using the cross-correlation of the \civ forest with quasars from the Sloan Digital Sky Survey (SDSS) data release 14 (DR14) \citep{sdssdr14}. \cite{gontcho2017} performed a similar study, using the SDSS-DR12 dataset \citep{sdssdr12} where they measured the effective optical depth $\bar\tau_c(z)$ and the linear bias parameters $\beta_c$ and $b_{Fc}$ of the \civ forest. The nature of the \civ doublet (as well as other metal doublets) allows for automated searches in quasar spectra to create catalogs that can be used in large-scale studies of metals in the universe \citep{anand2025, napolitano2023, anand2021}. Using such a technique, \cite{anand2025} and \cite{cooksey2013} found that the abundance of Carbon in the universe is larger at late times than at earlier times.

In this work, we present a BAO measurement from cross-correlations with the \civ forest using the 3-year main survey dataset from DESI DR2, collected between May 14th, 2021 and April 9th, 2024, using both quasars and ELGs as tracers. We begin in section \ref{sec:data} with a brief description of the datasets used in the analysis, including how we extract the \civ forest fluctuations from the quasar spectra. In section \ref{sec:method}, we present a summary of how we measure the correlations. Our method follows what was done in \cite{blomqvist2018} with updated analysis components following \lyabao. We also present the measured correlation functions from quasars and ELGs and discuss contamination from metals. In section \ref{sec:results} we present our main result and discuss the choices made for our baseline analysis. We also discuss validation tests on correlations with both the quasar and ELG samples. We conclude and suggest future studies in section \ref{sec:summary}.

\section{Data Sample}
\label{sec:data}

DESI is a ground-based Stage-IV dark energy experiment that began its main survey in 2021. Over the course of the 8-year survey, DESI will construct the largest 3D map of the Universe by collecting redshifts for over 60 million extragalactic sources. This map will allow for studies of the large-scale structure in the Universe and the effects of dark energy on the expansion of the universe. 

DESI is a multi-object fiber-fed spectroscopic instrument located at the Mayall Telescope at Kitt Peak National Observatory (KPNO) in the Tohono O'odham Nation in Arizona. Light from distant objects reflects off the 4-meter mirror into the optical corrector system and lands on the focal plane. The focal plane consists of 5,000 robotically controlled fiber positioners that simultaneously point at targets on the sky. Optical fibers route the light collected from the positioners to the spectrographs. There are 10 spectrographs, each with 3 channels (blue, red, and near-infrared), spanning wavelengths from 3600~\AA ~to 9800 \AA. The spectra are then processed by the spectroscopic pipeline \citep{guy-spectro}. More information about the DESI instrument and the spectrographs, the optical corrector, the focal plane, and the fiber system can be found in \cite{desi-inst2022}, \cite{miller-corrector}, \cite{silber-fp}, and \cite{poppett-fibers}, respectively. 

Over the course of the 8-year main survey, DESI will cover more than 14,000 square degrees of the sky. There are two programs that make up the main survey: dark and bright. The dark program targets galaxies (emission line galaxies and luminous red galaxies) and quasars when conditions are good, while the bright program targets bright galaxies and Milky Way stars when conditions are not good enough for the dark program \citep{schlafly-survey-ops}. 

In this paper, we use the DR2 dataset composed of data from the first three years of the DESI main survey. There are approximately twice as many quasars and nearly three times as many ELGs as there were in DR1.

\subsection{Quasar Sample}
\label{subsec:qso-sample}

\begin{figure}
    \centering
    \includegraphics[width=\linewidth]{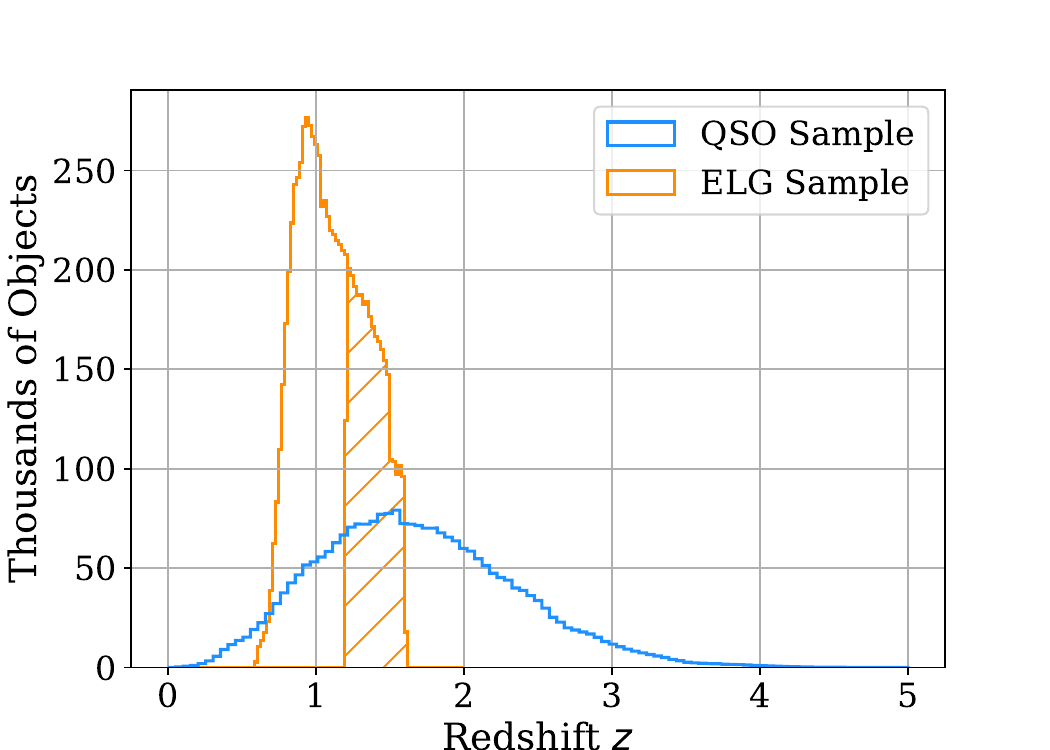}
    \caption{Redshift distributions of the quasar (blue) and the ELG catalogs (orange) used in this work. Only the ELGs with $z > 1.2$ will contribute to the cross-correlation and those redshifts are highlighted by the hatched histogram.}
    \label{fig:qso-z-dist}
\end{figure}

We use a modified version of the DESI DR2 quasar catalog used in \lyabao. The catalog is constructed based on results from the spectroscopic pipeline and the redshift fitting software \texttt{Redrock} \citep{bailey-redrock}. All targets are classified as either a galaxy, quasar, or star by \texttt{Redrock} and their redshift is measured. Quasar targets and objects classified as a quasar by \texttt{Redrock} are run through two additional afterburners. The first is the Mg~II afterburner \citep{chaussidon2023} that searches for broad Mg~II emission in quasar targets that \texttt{Redrock} identified as a galaxy. The second is \texttt{QuasarNet} \citep{green-qn, busca}, which uses a convolutional neural network to automatically classify quasar spectra.

The DR2 quasar catalog consists of 2,776,520 quasars. We modify this catalog and remove any quasars that are identified as Broad Absorption Line (BAL) quasars. BAL quasars have blueshifted absorption that is often coupled to \civ emission and these absorption troughs consequently affect the continuum in the \civ region. Though \cite{ennesser2022} and \cite{martini2025} showed that masking BAL features does not affect BAO measurements with the Lyman-$\alpha$ forest, quasars that have a BAL at the \civ emission line can have significant changes to the line profile that can lead to redshift errors. We identify BAL quasars with the BALFinder \citep{filbert2023, martini2025,guo-martini} and use the absorptivity index (AI) to select which BALs to exclude. We consider a quasar to be a BAL if it has AI $> 0$ and we exclude all BALs from our catalog. Our final quasar catalog consists of 2,498,218 quasars and the redshift distribution is shown in Figure~\ref{fig:qso-z-dist} as the blue histogram.

\subsection{Forest Samples}
\label{subsec:civ-forest-sample}

\begin{figure*}
    \centering
    \includegraphics[width=\linewidth]{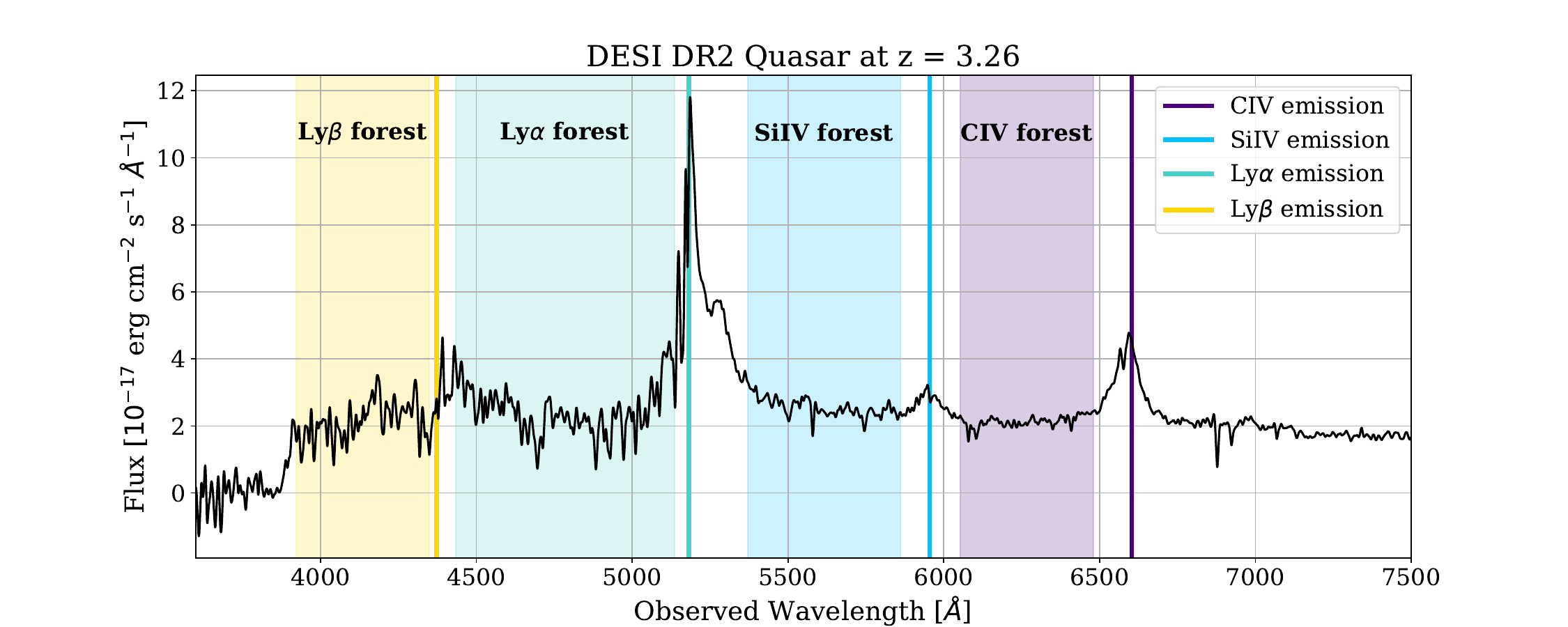}
    \caption{A smoothed quasar spectrum from the DESI DR2 dataset at $z=3.26$ (TargetID = 39627334130598165). The spectrum shows the emission peaks and forest regions for the two areas typically used in BAO analyses: Ly$\alpha$ (green) and Ly$\beta$ (yellow). In this analysis, we measure \civ absorption in the \civ region (purple) from 1420 to 1520 \AA~in the quasar rest-frame, and in the \siiv region (blue) from 1260 to 1375 \AA.}
    \label{fig:qso-spec}
\end{figure*}

We define the \civ region as the spectral region between 1420~\AA~$\leq \lambda_{\rm RF} <$ 1520~\AA~in rest frame wavelengths and we define the forest as the flux fluctuations due to \civ absorption in this region. This region avoids wavelengths close to the \civ and \siiv emission lines in the quasar spectra. Since quasar emission lines are broad, the wings of these lines can enter the forest region, especially \civ as it has a variable profile (e.g. \citealt{shen2016}). We also apply a cut for forests that fall outside of the observed wavelength range 3600 \AA~$\leq \lambda_{\rm OBS} < $ 7600 \AA. The lower limit is set by the DESI spectrographs and the upper limit corresponds to the middle of the transition region between the red and near-infrared channels. This effectively provides a quasar redshift cut of 1.36 $\leq z_{\rm q} \leq 4.35$. 

We also measure \civ flux fluctuations in the \siiv region, which we define as the spectral region between 1260~\AA~$ \leq \lambda_{\rm RF} < 1375$ \AA. The lower bound avoids the red wing of the Lyman-$\alpha$ and NV emission lines and the upper bound avoids the wing from the \siiv emission line. This corresponds to a quasar redshift cut of $1.6 \leq z \leq 5$. Figure \ref{fig:qso-spec} shows an example quasar spectrum from DR2 at $z=3.26$. The \civ and \siiv regions are highlighted, as well as the Lyman-$\alpha$ and Lyman-$\beta$ regions used in \lyabao. 

We use the average wavelength of the \civ doublet, $\lambda_{\rm c} = 1549.06$ \AA, to convert from spectral pixels to redshifts in each of the \civ and \siiv regions. 
We require each spectrum to have a number of valid pixels in the forest region with an integrated length larger than 25\% of the total length of the forest region. For the \civ region, this corresponds to 90 pixels and 110 pixels for the \siiv region. The pixel size is 0.8~\AA~(in the observer frame). 

We mask areas of the spectra that are prone to contamination from sky lines and absorption lines for Calcium and Sodium present in the interstellar medium. Because we removed BAL quasars from the catalog, we do not need to mask BAL regions in either forest region. \cite{martini2025} describes the updated strategy for masking BAL regions in the DR2 Lyman-$\alpha$ forest analysis. We discuss in section \ref{subsec:validation-tests} the impact that this strategy has on our results when we include the BAL quasars. Unlike the DR2 Lyman-$\alpha$ analysis, we do not have to account for Damped Lyman-$\alpha$ Absorption (DLA) systems in this work. DLAs occur blueward of the Lyman-$\alpha$ emission line and will therefore not appear in the \civ and \siiv regions.

In order to measure flux fluctuations in the \civ region, we measure the flux transmission field $\delta_{\rm q}$ for each quasar $q$ as a function of observed wavelength:
\begin{equation}
    \delta_{\rm q}(\lambda) = \frac{f_{\rm q}(\lambda)}{\overline{F}(\lambda)C_{\rm q}(\lambda)}-1,
\end{equation}
where $f_{\rm q}(\lambda)$ is the observed flux, $C_{\rm q}(\lambda)$ is the unabsorbed quasar continuum, and $\overline{F}(\lambda)$ is the mean transmission. The method for estimating the unabsorbed continuum is described in detail in \cite{cesar2023} and is performed with the \texttt{picca} software \citep{picca}. In short, the continuum fitting process consists of obtaining the mean expected flux, $\overline{F}(\lambda)C_{\rm q}(\lambda)$, over several iterations. To simplify the process, the mean expected flux is assumed to be a universal function of rest-frame wavelength with a first-degree polynomial correction in $\text{log}~\lambda$:
\begin{equation}
    \overline{F}(\lambda)C_{\rm q}(\lambda) = \overline{C}(\lambda_{\rm RF})\left(a_{\rm q} + b_{\rm q}\frac{\Lambda- \Lambda_{\rm min}}{\Lambda_{\rm max} - \Lambda_{\rm min}}\right),
\end{equation}
where $\Lambda \equiv \text{log} \lambda$. This correction accounts for differences in quasar luminosity and spectral diversity as well as the redshift evolution of $\overline{F}$. This iterative process solves for $\overline{C}(\lambda_{\rm RF})$, $a_{\rm q}$, and $b_{\rm q}$ and also takes into account the pipeline noise and applies a correction to account for the intrinsic variance of each forest region. There is an additional parameter, $\eta_{\rm LSS}$, that optimizes the weights assigned to each forest. For the Lyman-$\alpha$ forest, \cite{cesar2023} found that the optimal value for $\eta_{\rm LSS} = 7.5$. However, this value will change with different datasets, as it will depend on other properties that go into the weights calculation like the pipeline noise and the intrinsic variance in the particular forest region. Following the method in \cite{cesar2023}, we perform our own analysis of $\eta_{\rm LSS}$ for each of the \civ and \siiv regions and find $\eta_{\rm LSS}$ = 20 and $\eta_{\rm LSS}$ = 12 for the two forest regions, respectively.

After the above cuts are made and the continuum fitting process is complete, our \civ region sample contains 1,524,506 forests and the \siiv region sample has 1,130,354 forests.

\subsection{Emission Line Galaxy Sample}
Our sample of ELGs comes from the DESI DR2 large-scale structure (LSS) catalogs used in \galbao. The LSS catalog pipeline is detailed in \cite{lss-cat} (section 7.1). The ELG sample is split into two redshift bins spanning $0.8 < z < 1.1$ and $1.1 < z < 1.6$, containing 2.7 million and 3.8 million ELGs, respectively (\galbao). In this work, we use the clustering catalog for the full ELG sample. The clustering catalogs are the same as the full LSS catalogs; however, they remove extra columns and apply a cut to select only good spectroscopic observations. A good observation is one that has no Redrock \texttt{ZWARN} bits set \citep{desi-edr}.

The ELG catalog consists of redshifts between $0.6 < z < 1.6$, though any object with a redshift less than $z \sim 1.2$ will not be included in the correlations as the comoving separation will be larger than the range we study. The distribution of redshifts is shown in Figure \ref{fig:qso-z-dist} as the orange histogram. The hatched histogram shows the objects with $z > 1.2$. We do not apply any other cuts to the catalog, and our final ELG sample contains 3.1 million objects.

\section{Analysis}
\label{sec:method}

\begin{figure}[t]
    \centering
    \includegraphics[width=\linewidth]{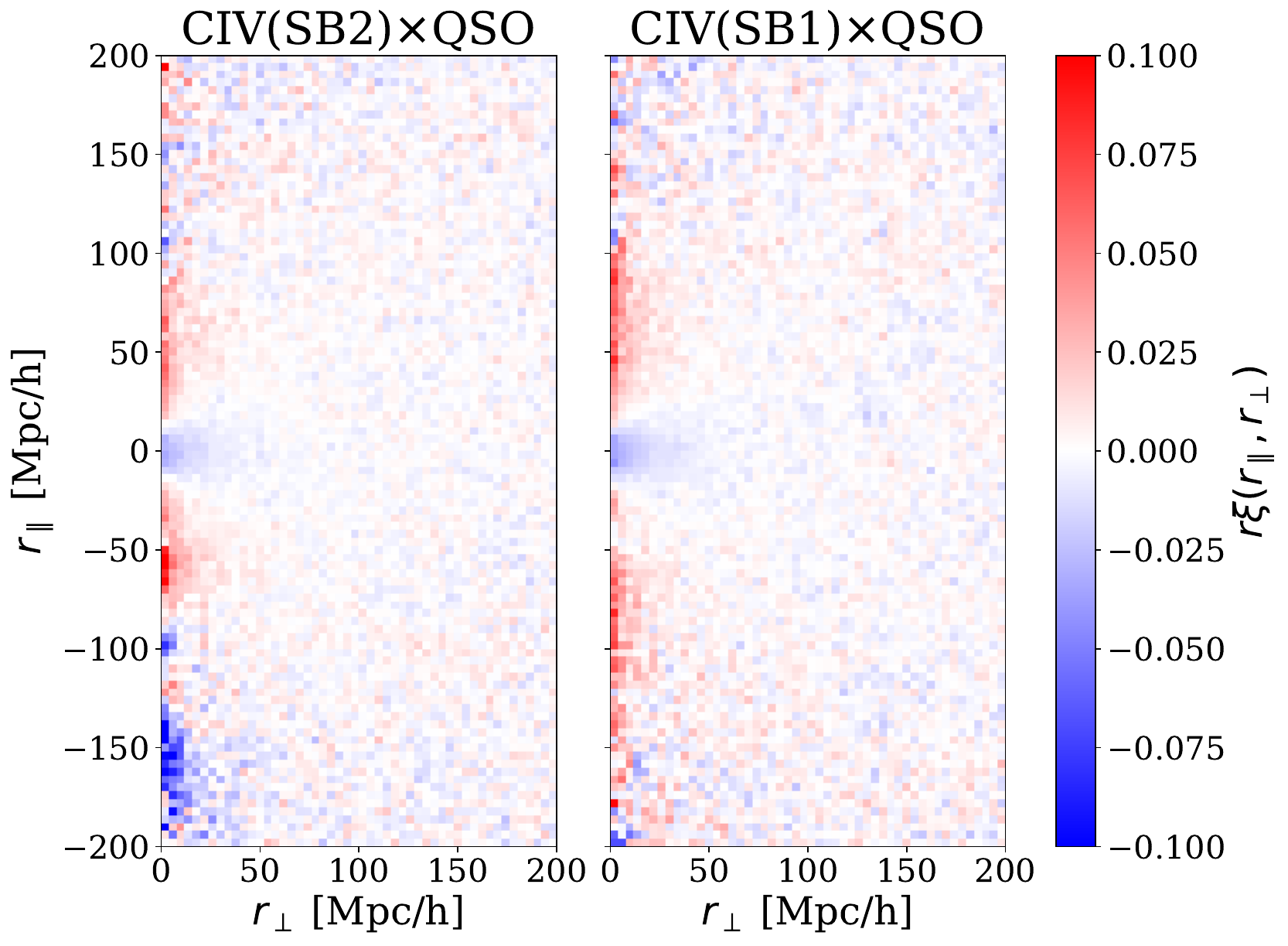}
    \caption{Two-dimensional correlation function shown as \rp and \rt for the two correlations: CIV(SB2)$\times$QSO and CIV(SB1)$\times$QSO. The BAO feature is not easily seen by eye but is observed as a half ring at $r \sim 100$\mpc. Both correlations are multiplied by $r$ for visualization purposes. 
    There are oscillatory features at $r_\parallel < 0$ in CIV(SB2) caused by quasar redshift errors (discussed in section \ref{subsec:fit}).}
    \label{fig:2d-corr-qso}
\end{figure}

We measure the cross-correlation function with \civ absorption in two regions: the \civ region and the \siiv region, which we refer to as side-band 2 (SB2) and side-band 1 (SB1), respectively, throughout the rest of this paper. We measure the cross-correlation in each side-band using both quasars and ELGs as tracers. We follow the methods used in previous analyses to measure the correlation functions, calculate the distortion and covariance matrices, and model the contamination due to metals. The correlation and covariance calculations are the same as in \cite{desi-dr1-kp6}, but improvements have since been made to the computation of the distortion and metal matrices. These improvements are described in Section 3 of \lyabao. We calculate the correlations and the distortion matrix with \texttt{picca}\footnote[1]{\texttt{picca}: \url{https://github.com/igmhub/picca/tree/v9.13.0}}. We follow the same method as \lyabao to calculate the covariance. The metal matrices are calculated with \texttt{Vega} \citep{cuceu2023,dmdb2020} during the fit. 

\subsection{Correlations}
\label{subsec:correlations}

We calculate the correlation function by binning in comoving coordinates along (\rp) and across (\rt) the line of sight, extending out to 200 \mpc in both positive and negative \rp and positive \rt. Because the cross-correlation is not symmetric and we need to extend to negative separations in \rp, we define a positive separation along the line of sight when the quasar or ELG is in front of the \civ forest pixel. To calculate separations in comoving coordinates, we assume the $\Lambda$CDM cosmology from \cite{planck2018}, hereafter ``Planck2018'', as our fiducial cosmology. This is the same fiducial cosmology used in \lyabao.

\begin{figure}
    \centering
    \includegraphics[width=\linewidth]{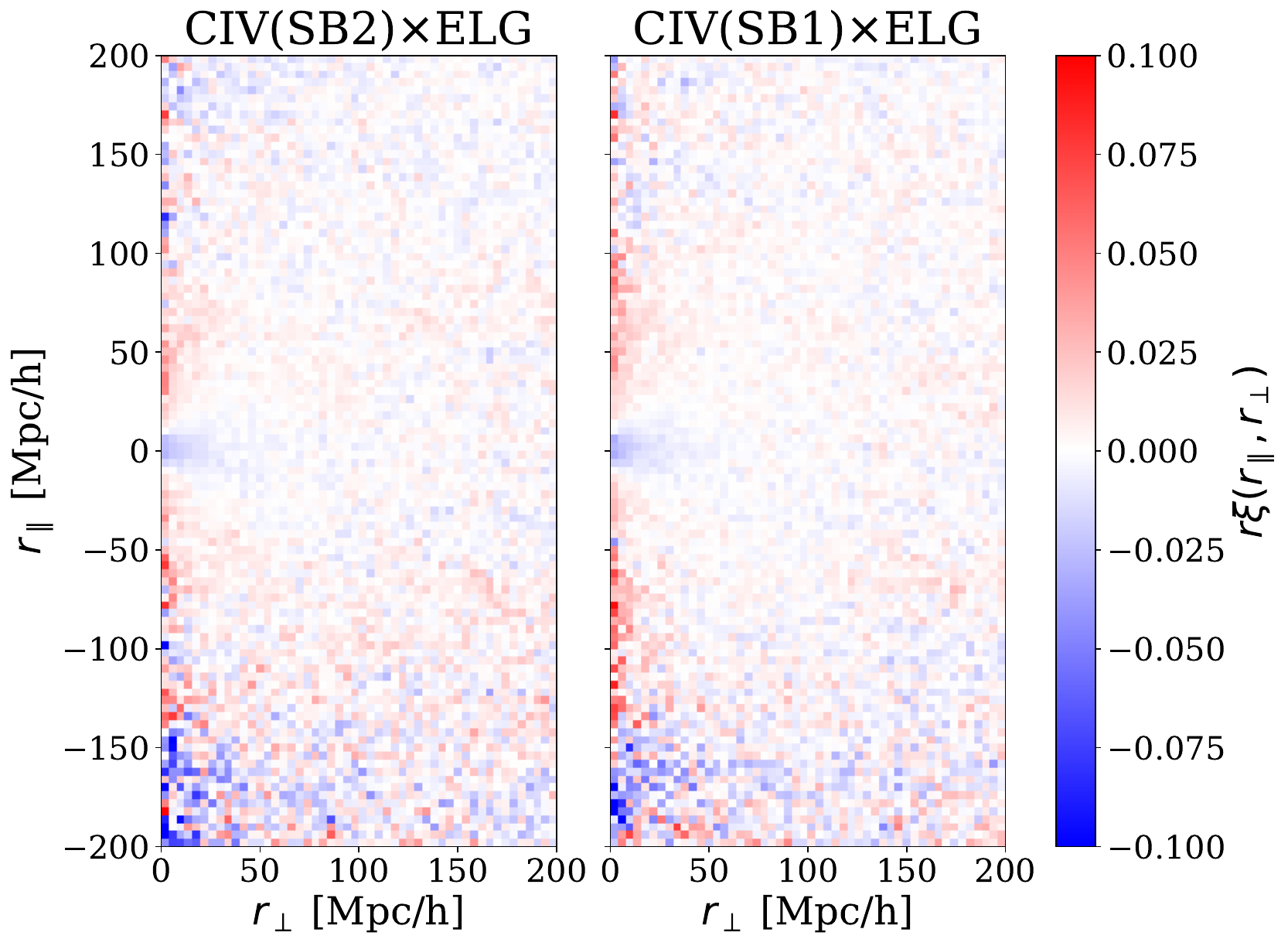}
    \caption{Two-dimensional correlation function shown as \rp and \rt for the two correlations: CIV(SB2)$\times$ELG and CIV(SB1)$\times$ELG. The BAO feature is not easily seen by eye but is observed as a half ring at $r \sim 100$\mpc. Both correlations are multiplied by $r$ for visualization purposes. 
    The data is noisy so it is difficult to see any oscillatory features at $r_\parallel < 0$ in CIV(SB2).}
    \label{fig:2d-corr-elg}
\end{figure}

We show the measured two-dimensional correlations in \rp and \rt with quasars and ELGs for each side-band in Figures \ref{fig:2d-corr-qso} and \ref{fig:2d-corr-elg}. We multiply the correlation by the total separation $r$ and center the colorbar around zero for visualization purposes. The BAO can usually be seen as a half ring at $r \sim 100$\mpc, though the data here is noisy and it is harder to see. In SB2 in Figure \ref{fig:2d-corr-qso}, there are oscillatory features at small \rt for negative \rp that are not present in SB1. We believe these oscillations are due to quasar redshift errors \citep{gordon2025, youles2022} and we discuss how to mitigate them in section \ref{subsec:fit}. These errors are also present in the ELG correlation in Figure \ref{fig:2d-corr-elg}, but the data is noisier and it is not as easily seen in the figure.

\begin{figure}
    \centering
    \includegraphics[width=\linewidth]{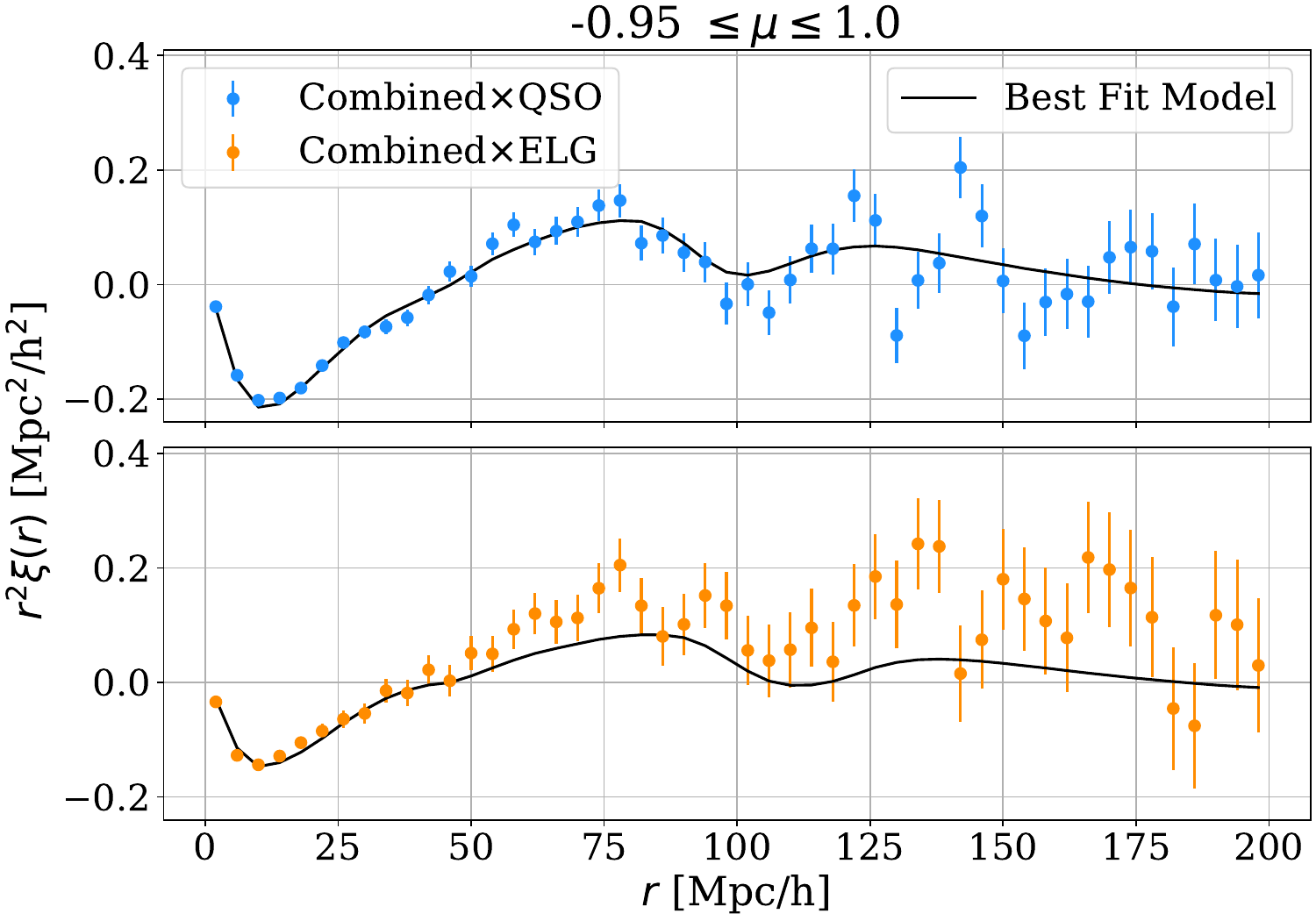}
    \caption{Weighted one-dimensional correlation averaged over bins of total separation $r$ and lines of sight $\mu$ when using quasars (blue) and ELGs (orange) as tracers. For each tracer, we are showing the correlation when combining the two side-bands, and we compress the two-dimensional data and best-fit model into one wedge between $-0.95 \leq \mu \leq 1.0$. We multiply both correlations by $r^2$ for visualization purposes. The BAO ``peak'' is seen as a trough near 100\mpc for both tracers.}
    \label{fig:wedges}
\end{figure}

Figure \ref{fig:wedges} shows the two-dimensional correlation of Figures \ref{fig:2d-corr-qso} and \ref{fig:2d-corr-elg} reduced to a weighted one-dimensional correlation that is averaged over bins of total separation $r = (r_\parallel^2 + r_\perp^2)^{1/2}$. The wedges are defined by a range of the cosine between $r$ and the line of sight, $\mu = r_\parallel/r$. $|\mu| = 1$ is along the line of sight while $\mu = 0$ is across the line of sight. In this case, we are compressing the data and best fit model (black line) from both SB2 and SB1 into one wedge between $-0.95 \leq \mu \leq 1.0$ for quasars (blue points) and ELGs (orange points). These wedges are not used for fitting the data. The BAO feature can be seen as a trough near 100\mpc.

\subsection{Contamination from Other Transition Lines}
\label{subsec:metals}

\begin{figure*}
    \centering
    \includegraphics[width=0.8\linewidth]{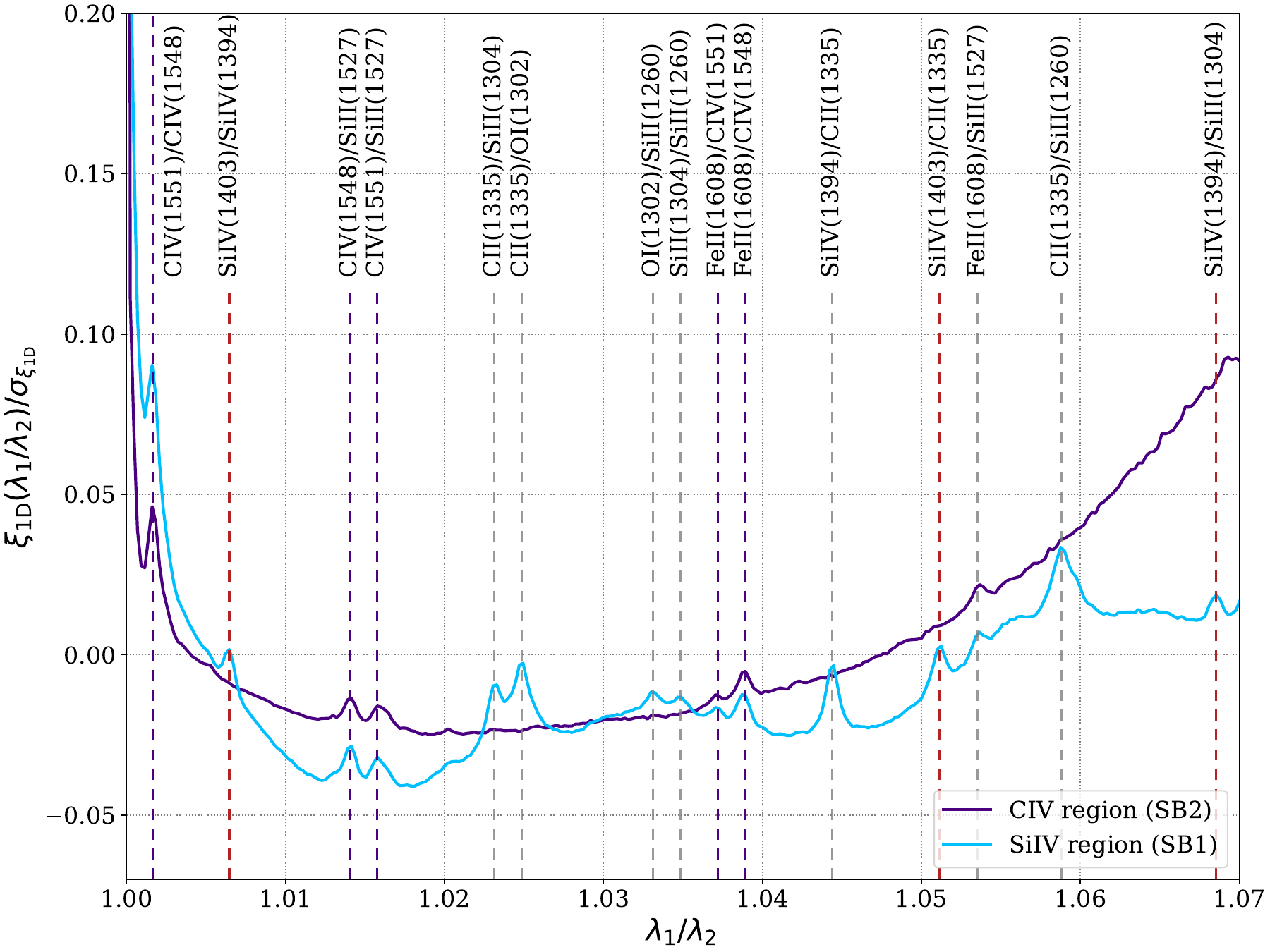}
    \caption{One-dimensional auto-correlation function $\xi_{\rm 1D}$ as a function of wavelength ratio $\lambda_1/\lambda_2$ for pixel pairs in SB2 (indigo) and SB1 (blue), ordered such that $\lambda_1 > \lambda_2$. Vertical dashed lines indicate prominent correlations involving the \civ doublet (indigo), the \siiv doublet (red), and other metal-metal transitions (grey). The metals involving the \civ doublet are the only lines that we include in our correlations. These are FeII(1608) and SiII(1527).}
    \label{fig:metals1}
\end{figure*}

There are small contributions from other absorption by metals in the intergalactic medium (IGM) that can produce additional signal in the cross-correlations. Historically, ``metals`` refers to any element that is not neutral hydrogen, meaning that \civ itself would be classified as a metal. In this work, we define metal absorption as absorption caused by transitions other than \civ.

Metal contamination in the Lyman-$\alpha$ forest has been extensively studied \citep{guy2025, yang2022,bautista2017,pieri2014b}, however, it has been studied less so in the \civ and \siiv regions. Following \cite{blomqvist2018}, we identify metal contamination in the \civ and \siiv regions by measuring the one-dimensional auto-correlation function, $\xi_{1{\rm D}}$, using the forest samples described in section \ref{subsec:civ-forest-sample}. Figure \ref{fig:metals1} shows $\xi_{1{\rm D}}$ as a function of wavelength ratio $\lambda_1/\lambda_2$ for the \civ region (SB2, indigo) and the \siiv region (SB1, blue).

There are prominent peaks from correlations between the \civ and \siiv doublets, \civ- and \siiv-metal, and metal-metal pairs. Since we are measuring correlations using \civ as the absorber, only the peaks associated with either line of the \civ doublet will produce contributions to the measured correlations. These lines are SiII(1527) ($\lambda = 1526.71$~\AA) and FeII(1608) ($\lambda = 1608.45$ \AA) and are highlighted with purple dashed lines in Figure~\ref{fig:metals1}.

The contributions from the two metal lines will appear in both the QSO and ELG cross-correlations, as they are dependent on the absorber (\civ) and not the tracer. The maximum contributions from the different metals occur at different redshifts along the line of sight at \rt = 0 and $r_\parallel \approx \frac{c}{H(z)}(1+z)(\lambda_{\rm m} - \lambda_{\rm CIV})/\lambda_{\rm CIV}$. The \rp values for the \civ-metal transition are given in Table \ref{table:metals}.

\begin{table*}[]
    \centering
    \begin{tabular}{ccccc} \hline \hline 
        Absorber pair & $\lambda_1$[\AA] & $\lambda_2$[\AA] & $\lambda_1/\lambda_2$ & \rp[\mpc] \\ \hline 
        C\,{\sc iv}(1548)-Si\,{\sc ii}(1527) & 1548.20 & 1526.71 & 1.014 & -41.22 \\
        C\,{\sc iv}(1551)-Si\,{\sc ii}(1527) & 1550.78 & 1526.71 & 1.016 & -46.08 \\
        Fe\,{\sc ii}(1608)-C\,{\sc iv}(1548) & 1608.45 & 1548.20 & 1.039 & 115.53 \\
        Fe\,{\sc ii}(1608)-C\,{\sc iv}(1551) & 1608.45 & 1550.78 & 1.037 & 110.41 \\ \hline

    \end{tabular}
    \caption{Metal correlations involving the \civ doublet lines for the two metal species identified in Figure \ref{fig:metals1} that will impact the CIV(SB2) and CIV(SB1) correlations. The second and third columns list the rest-frame wavelength of the absorber-metal pair. The fourth column gives the wavelength ratio of the two lines. The final column gives the line-of-sight separation calculated at $z=2$ for the quasar measurement. These metal correlations will occur at slightly different separations (using $z=1.5$) for the ELG measurements.  }
    \label{table:metals}
    
\end{table*}

\section{Results}
\label{sec:results}

The model used to measure the BAO scale has been well described and widely studied in previous works (\cite{dmdb2020, gordon2023, bault2024,desi-dr1-kp6, desi-dr2-bao}; etc.). We follow the same prescription here and discuss key differences in this work in section \ref{subsec:baseline}. We use the \texttt{Vega}\footnote[2]{\texttt{Vega}: \url{https://github.com/andreicuceu/vega} version 1.3.1} package to model the correlations and metals and for parameter inference.

The BAO parameters are defined as:
\begin{equation}
    \label{eq:ap_at}
     \alpha_\parallel = \frac{[D_{\rm H}(z_{\rm eff})/r_{\rm d}]}{[D_{\rm H}(z_{\rm eff})/r_{\rm d}]_{\rm fid}} \text{~ and~} \alpha_\perp = \frac{[D_{\rm M}(z_{\rm eff})/r_{\rm d}]}{[D_{\rm M}(z_{\rm eff})/r_{\rm d}]_{\rm fid}},
\end{equation}
where $z_{\rm eff}$ is the effective redshift of the measurement and ``fid'' represents the Planck2018 fiducial cosmology used in this work where $\alpha_\parallel = \alpha_\perp = 1$. $D_{\rm H}(z)$ is the Hubble distance and is defined as
\begin{equation}
    \label{eq:hubble}
    D_{\rm H}(z) = \frac{c}{H(z)},
\end{equation}
and $D_{\rm M}(z)$ is the transverse comoving distance. For a flat universe, $D_{\rm M}$ is defined as
\begin{equation}
    \label{eq:dm}
    D_{\rm M}(z) = \frac{c}{H_0}\int_0^z \frac{dz'}{H(z')/H_0}.
\end{equation}

When there is not enough signal to measure BAO from $\alpha_\parallel$ and $\alpha_\perp$, we can instead measure BAO from the isotropic signal: 
\begin{equation}
    \label{eq:a_iso}
    \alpha_{\rm iso} = \frac{[D_{\rm V}(z_{\rm eff})/r_{\rm d}]}{[D_{\rm V}(z_{\rm eff})/r_{\rm d}]_{\rm fid}},
\end{equation}
where $D_{\rm V}(z)$ is the spherically-averaged distance over the radial and transverse directions and is defined as:
\begin{equation}
    \label{eq:dv}
    D_{\rm V}(z) \equiv [zD_{\rm M}(z)^2D_{\rm H}(z)]^{1/3}
\end{equation}
In equations \ref{eq:ap_at} and \ref{eq:a_iso}, $r_{\rm d}$ is the sound horizon at the drag epoch. The distances $D_{\rm H}$, $D_{\rm M}$, and $D_{\rm V}$ are inferred relative to $r_{\rm d}$, the constrained quantities are the ratios $D_{\rm H}/r_{\rm d}$, $D_{\rm M}/r_{\rm d}$, or $D_{\rm V}/r_{\rm d}$. In all results discussed in this paper, we are measuring isotropic BAO unless otherwise specified.

\subsection{Baseline Fit}
\label{subsec:baseline}
We present results for BAO measurements with two separate tracers: one with quasars and one with~ELGs. The two quasar cross-correlations in the baseline fit include \civ absorption in SB2 (CIV(SB2)$\times$QSO) and in SB1 (CIV(SB1)$\times$QSO). Similarly for ELGs, our baseline fit contains the~two cross-correlations with \civ absorption in SB2 (CIV(SB2)$\times$ELG) and in SB1 (CIV(SB1)$\times$ELG). We also include the covariance matrix to account for small cross-covariance between the two correlation functions for each tracer. Following \lyabao, we fit all correlations over the range $30 < r < 180$\mpc. For both the quasar and ELG measurements, we exclude the wedge region along the line of sight at negative \rp (between $-1.0 \leq \mu < -0.95$) in the SB2 region only. We discuss why later in this section. 

Our baseline fit contains 7 free parameters: the BAO parameter $\alpha_{\rm iso}$ (8 free parameters when fitting for $\alpha_\parallel$ and $\alpha_\perp$), the bias and redshift space distortion (RSD) parameters for the \civ forest: $b_{\rm CIV(eff)}$ and $\beta_{\rm CIV(eff)}$, the bias parameters for the two metal species identified in section \ref{subsec:metals}: $b_{\rm FeII(1608)}$ and $b_{\rm SiII(1527)}$, and two parameters that describe redshift errors: $\sigma_{\rm v}$ and $\Delta r_\parallel$, which account for peculiar velocities and systematic shifts in redshifts \citep{bault2024} for each tracer. We fix $\beta_{\rm FeII(1608)}$ and $\beta_{\rm SiII(1527)}$ to 0.5, as done in many previous analyses (including \cite{blomqvist2018}, \cite{dmdb2020}, \cite{desi-dr1-kp6}, and \lyabao) and based on the results from \cite{fontribera2012}. The SiII(1527) metal line appears strongest in the correlations at (\rt,\rp) = (0, $\sim$-45)\mpc. This point is in the wedge region that is excluded from SB2 in the fit, and therefore $b_{\rm SiII(1527)}$ is poorly constrained. We mitigate this by adding a Gaussian prior using the best-fit value from the CIV(SB1) fits for each tracer as the $-1.0 \leq \mu < -0.95$ wedge is not masked in this region. Because \civ is a doublet, there are additional contributions to the $\sigma_{\rm v}$ parameter that cause it to be poorly constrained. We therefore add a Gaussian prior to $\sigma_{\rm v}$ using the value from \lyabao. Future work should look to improve on the $\sigma_{\rm v}$ prior for the ELG measurements. ELG redshifts in DESI are more precise as the redshift is determined by the [O\,{\sc ii}] doublet. Inherently, this means that the $\sigma_{\rm v}$ value for ELGs would be smaller than that of quasars. However, given the contribution from the \civ doublet to $\sigma_{\rm v}$, we choose to use the same prior for both the quasar and ELG measurements.

We fix the quasar or ELG bias ($b_{\rm q}$ or $b_{\rm e}$), as it is degenerate with the bias of the \civ forest, to the value calculated at the effective redshift for each tracer. 
The redshift evolution of the tracer bias is given by
\begin{equation}
    \label{eq:tracer-bias}
    b_t(z) = b_t(z_{\rm eff})\left(\frac{1+z}{1+z_{\rm eff}}\right)^{\gamma - 1}.
\end{equation}
For quasars, $\gamma = 1.44$ \citep{dmdb2019} and for the Lyman-$\alpha$ forest $\gamma = 2.9$ \citep{mcdonald2006}. We estimate $\gamma_{\rm CIV} = 1.5$ by measuring $b_{\rm CIV}$ at $z_{\rm eff} = 1.92$ and $z_{\rm eff}=1.47$ and fitting for $\gamma$ in equation \ref{eq:tracer-bias}. For ELGs, we used equation C.2 from \cite{chaussidon2025} (with $a = 0.15$ and $b=0.59$) and fit it with equation \ref{eq:tracer-bias} between $1.0 < z < 2.5$ and found $\gamma_{\rm ELG} = 1.3$.

\begin{figure}
    \centering
    \includegraphics[width=\linewidth]{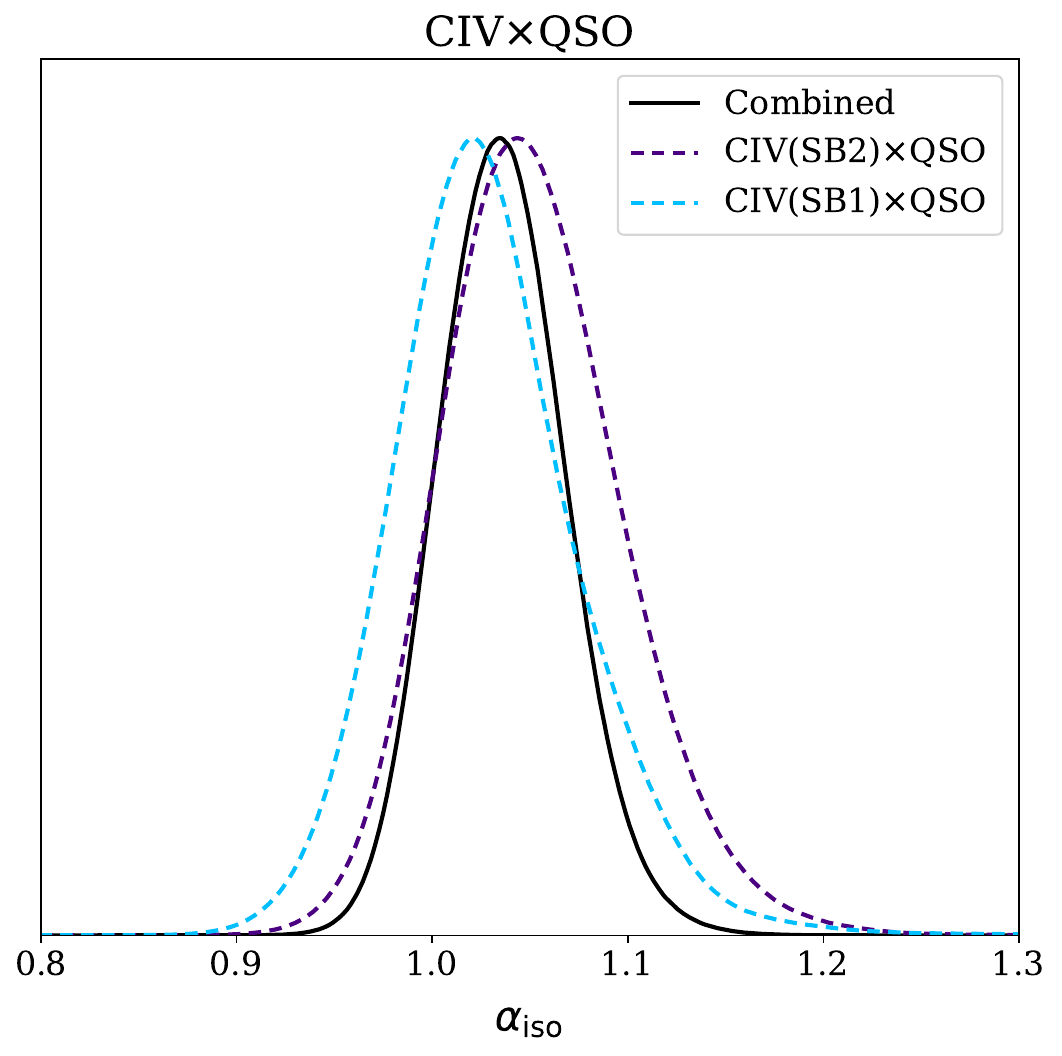}
    \caption{Likelihood of $\alpha_{\rm iso}$ from the combined fit of the CIV$\times$QSO cross-correlation (solid black line) as well as for fits to the individual CIV(SB2)$\times$QSO (purple dashed line) and CIV(SB1)$\times$QSO (blue dashed line) side bands.}
    \label{fig:bao-results-qso}
\end{figure}

We present a result of the BAO measurement using the isotropic signal. Our best fit value of $\alpha_{\rm iso}$ from the combined fit with quasars is:
\begin{equation}
    \alpha_{\rm iso} = 1.031 \pm 0.030,
\end{equation}
with a reduced $\chi^2$ of $5808.76/(5891-7)=0.987$ and a fit probability of 75.5\%. We show the likelihood on $\alpha_{\rm iso}$ for the combined fit, as well as the individual side band fits, in Figure \ref{fig:bao-results-qso}. We quantify the detection of this BAO measurement by fitting a model with no BAO for the combined fit. We find a 4.2$\sigma$ detection of the BAO signal from the measurement with quasars. 

\begin{figure}
    \centering
    \includegraphics[width=\linewidth]{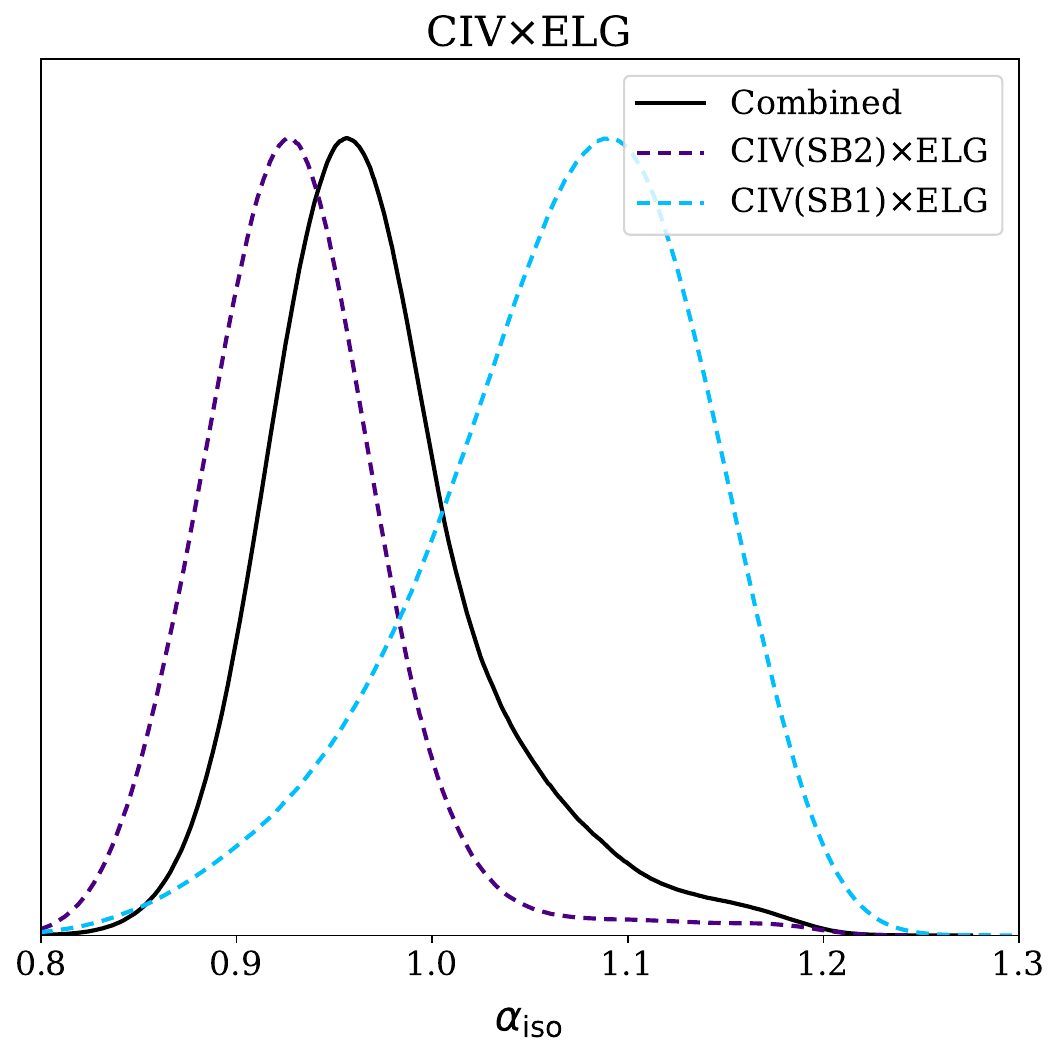}
    \caption{Likelihood of $\alpha_{\rm iso}$ from the combined fit of the CIV$\times$ELG cross-correlation (solid black line) as well as for fits to the individual CIV(SB2)$\times$ELG (purple dashed line) and CIV(SB1)$\times$ELG (blue dashed line) side bands.}
    \label{fig:bao-results-elg}
\end{figure}

For ELGs, the best fit value for isotropic BAO from the combined fit is:
\begin{equation}
    \alpha_{\rm iso} = 0.953 \pm 0.038,
\end{equation}
with a reduced $\chi^2$ of $6116.76/(5891-7)=1.039$ and a fit probability of 1.7\%. We show the likelihood of $\alpha_{\rm iso}$ for the combined fit with ELGs and the two side bands in Figure \ref{fig:bao-results-elg}. We measure BAO at 2.5$\sigma$ significance from the measurement with ELGs. 

\cite{boryana2025} and \cite{debelsunce2024} have investigated systematic uncertainties on the BAO shift from Lyman-$\alpha$ forest measurements when using linear theory. Following these results, \lyabao added a systematic uncertainty of $\Delta\alpha_\parallel$=0.3\%,$\Delta\alpha_\perp$=0.3\% to the covariance matrix. We expect the systematic uncertainty for non-linear clustering in this work to be similar to 0.3\%; however, our errors are 10 times larger at 3\% so we do not include a systematic uncertainty in our analysis. We also perform numerous validation tests, discussed in Section \ref{subsec:validation-tests}, that do not indicate any other source of significant systematic uncertainties.

\begin{table*}[t]
    \centering

    \resizebox{\textwidth}{!}{
    \hspace{-2.6cm}
    \begin{tabular}{cc|ccc|ccc} 
    
    \hline \hline 
        \multirow{2}{*}{Parameter} & \multirow{2}{*}{Priors} &  & Best fit - QSO & & & Best fit - ELG \\ 
                                   &                         & CIV(SB2) & CIV(SB1) & Combined  & CIV(SB2) & CIV(SB1)       & Combined \\ 
        \hline 
        $\alpha_{\rm iso}$                 & $\mathcal{U}$[0.5, 1.5] & $1.05 \pm 0.05$ & $1.02 \pm 0.04$ & $1.032 \pm 0.030$ & $0.92 \pm 0.04$ & $1.09 \pm 0.06$ & $0.95 \pm 0.04$ \\
        
        $b_{\rm CIV_{eff}}$                & $\mathcal{U}$[-2.0, 0.0] & $-0.0115 \pm 0.0008$ & $-0.0133 \pm 0.0006$ & $-0.0129 \pm 0.0005$ & $-0.0131 \pm 0.0014$  & $-0.0124 \pm 0.0008$ & $-0.0123 \pm 0.0007$ \\
        
        $\beta_{\rm CIV_{eff}}$            & $\mathcal{U}$[0.0, 4.0] & $0.59 \pm 0.14$   & $0.45 \pm 0.09$ & $0.46 \pm 0.07$ & $0.25 \pm 0.18$ & $0.20 \pm 0.13$ & $0.24 \pm 0.11$ \\
        
        $10^3b_{\rm FeII(1608)}$           & $\mathcal{U}$[-0.2, 0.0] & $-0.38 \pm 0.20$ & $-0.56 \pm 0.20$ & $-0.47 \pm 0.14$ & $-0.59 \pm 0.20$ & $-0.41 \pm 0.18$ & $-0.41 \pm 0.14$ \\
        
        \multirow{2}{*}{$10^3b_{\rm SiII(1527)}$} & $\mathcal{N}$[-1.49, 0.18] & $-1.23 \pm 0.17$  
        & $-1.50 \pm 0.18$ & $-1.32 \pm 0.12$ & - & - & - \\ 
        & $\mathcal{N}$[-1.60, 0.27] & - & - & -&  $-1.51 \pm 0.27$ & $-1.60 \pm 0.28$ & $-1.51 \pm 0.19$ \\
        
        $\sigma_{\rm v}$($h^{-1}$ Mpc)     & $\mathcal{N}$[3.18, 0.5] & $3.2 \pm 0.5$ & $3.4 \pm 0.5$ & $3.3 \pm 0.5$ & $3.1 \pm 0.5$ & $3.1 \pm 0.5$ & $3.0 \pm 0.5$ \\
        
        $\Delta r_\parallel$($h^{-1}$ Mpc) & $\mathcal{U}$[-5.0, 5.0]  & $-0.9 \pm 1.0$ & $0.0 \pm 0.7$ & $0.1 \pm 0.6$ & $3.4 \pm 1.1$ & $-0.3 \pm 0.6$ & $0.5 \pm 0.6$ \\ 
        
        \hline
        
        $z_{\rm eff}$     & - & 1.97 & 1.87 & 1.92 & 1.47 & 1.47 & 1.47 \\
        
        $\chi^2_{\rm min}$ & - & 2684.60 & 3109.73 & 5808.76 & 2909.94 & 3193.43 & 6116.76 \\
        
        $N_{\rm bin}$     & - & 2789 & 3102 & 5891 & 2789 & 3102 & 5891 \\
        
        $N_{\rm param}$   & - & 7 & 7 & 7 & 7 & 7 & 7 \\
        probability       & - & 0.91 & 0.42 & 0.75 & 0.04 & 0.11 & 0.01 \\ \hline 
        
    \end{tabular}
    }
    \caption{Best-fit baseline results for CIV(SB2), CIV(SB1), and the combined fit for both the QSO and ELG tracers when fitting for isotropic BAO. The first column lists the parameters in the fit and the second column lists the priors used. Separate priors are used on $b_{\rm SiII(1527)}$ when fitting with quasars or ELGs. The rest of the columns give the best-fit results for each parameter depending on which tracer is used. The first three columns give results for quasars while the last three columns give results for ELGs. The bottom section of the table gives the effective redshift, $\chi^2$, the number of data points in the fit, the number of free parameters, and the fit probability.}
    \label{table:baseline-fit}
\end{table*}

The results for the best fit parameters for the combined fits and fits to CIV(SB2) and CIV(SB1) for the two tracers, as well as the priors used, are given in Table \ref{table:baseline-fit}. All parameters have a flat prior except for $\sigma_{\rm v}$, which has a Gaussian prior given by values from the combined fit from \lyabao, and $b_{\rm SiII(1527)}$, which has a Gaussian prior for the combined fit and the fit to SB2 given by the best-fit value from SB1. We show the agreement between parameters for the combined and two individual fits for the two tracers in Figures \ref{fig:triangle-baseline-qso} and \ref{fig:triangle-baseline-elg} in Appendix \ref{sec:app_b}. For both measurements, all parameters show a good agreement between the three fits. 

Our results from the CIV$\times$QSO measurement are in good agreement with the results from \cite{blomqvist2018}. They measured BAO to 6\% precision at a significance of 1.7$\sigma$. We have improved the uncertainty of $\alpha_{\rm iso}$ by a factor of 2 and have increased the significance of the detection by over 2$\sigma$. Our results for $\sigma_{\rm v}$ and $\Delta r_\parallel$ differ, however, this is not unexpected as the templates and method used to estimate quasar redshifts have improved \citep{bault2024,brodzeller2023}.

\subsection{Goodness of Fit}
\label{subsec:fit}

\begin{figure}
    \centering
    \includegraphics[width=\linewidth]{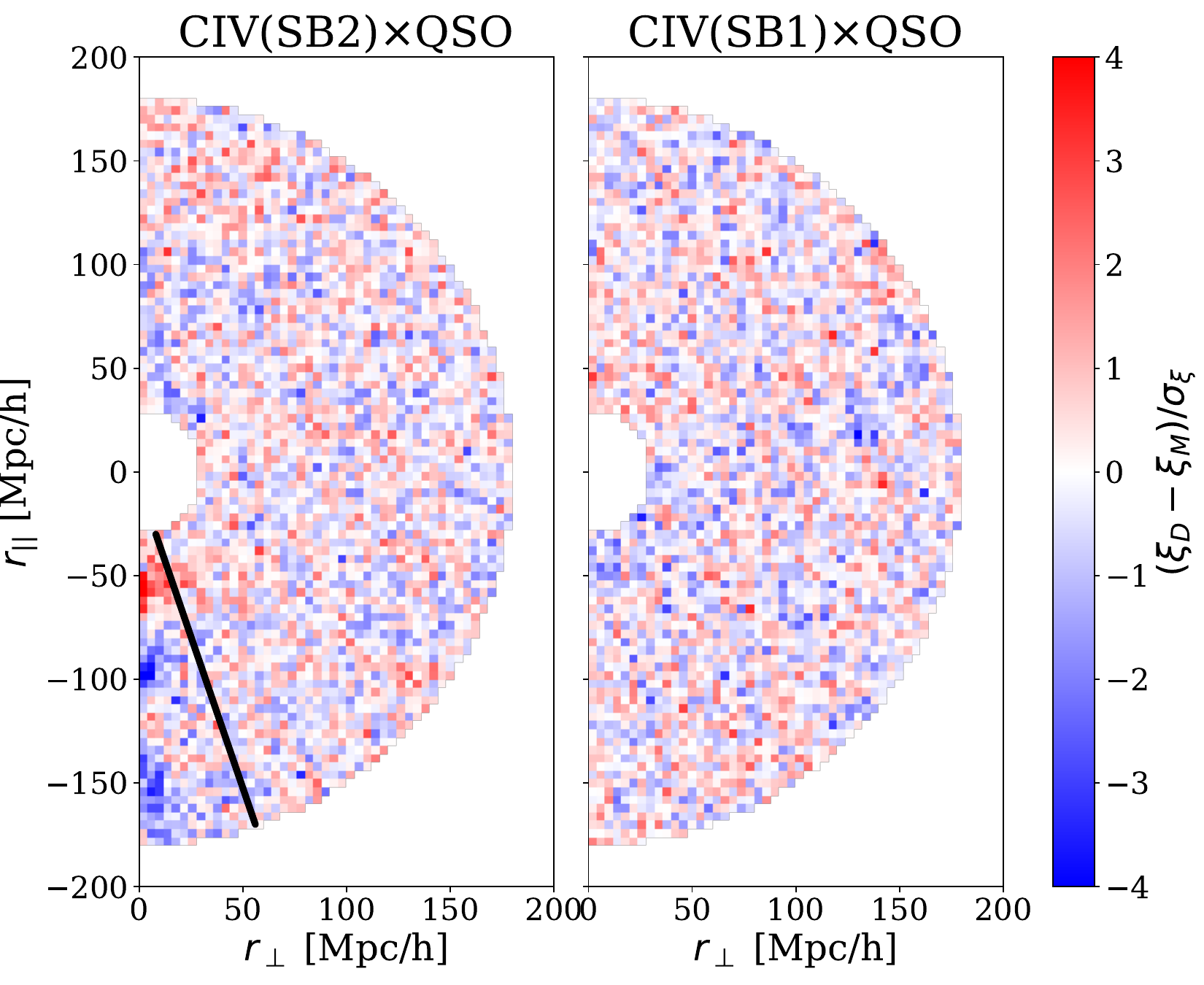}
    \caption{Residuals of the best-fit model for the baseline analysis with data for the CIV$\times$QSO cross-correlation shown for each side band. The area to the left of the black line is the wedge ($-1.0 \leq \mu < -0.95$) excluded from the fit due to redshift errors, where the oscillations are clearly present. SB1 has no visible oscillations because they occur at $|r_\parallel| > 200$\mpc.}
    \label{fig:residuals}
\end{figure}

We explore further the fit to the quasar measurement by studying the normalized residuals of the best fit model and the measured correlations, $(\xi_{\rm D} - \xi_{\rm M})/\sigma_\xi$, for the two correlation functions where $\xi_{\rm D}$ is the measured correlation, $\xi_{\rm M}$ is the best fit model, and $\sigma_\xi$ is the uncertainty of the correlation and is defined as the square root of the diagonal of the covariance matrix. The two-dimensional residuals for each correlation are shown in Figure \ref{fig:residuals}. The area to the left of the black line is the wedge along the line of sight at negative \rp (between $-1.0 \leq \mu < -0.95$) that is excluded from the fit. 

For the quasar measurement, we find no significant disagreement between the data and the model in SB1. However, in SB2 we see oscillatory structure from $-180 < r_\parallel < -50$\mpc along the line of sight ($r_\perp \sim 0$\mpc). These oscillations are the result of quasar redshift errors producing spurious correlations during the continuum fitting process. The \rp where these correlations between quasar-pixel pairs in CIV(SB1) are strongest is larger than 200 \mpc and, therefore, are not visible in the right panel of Figure \ref{fig:residuals}. The oscillations are also present in the cross-correlation with ELGs, though the measurement is noisier and it is harder to see by eye. 

\cite{casas2025} showed a small bias on $\alpha_\parallel$ on the order of 0.5\% produced by spurious correlations from quasar redshift errors during the continuum fitting process. \cite{gordon2025} developed a method to remove this contamination. For quasar-pixel pairs in the cross-correlation, the contamination is strongest when the tracer quasar is closer to the host quasar of the forest. \cite{gordon2023} and \cite{casas2025} showed that removing close quasar-pixel pairs removes most of the spurious correlations and therefore reduces the small bias on $\alpha_\parallel$. We remove close pairs by performing cuts in the correlations to remove pairs that are spatially close and that are close in redshift. Shown in Figure \ref{fig:variations}, removing the close pairs does not affect the results on $\alpha_{\rm iso}$. 

With our exclusion of the wedge along the line of sight at negative \rp in SB2 (between $-1.0 \leq \mu < -0.95$), we find that we have a good agreement between the data and the best-fit model.

The best fit for CIV(SB1)$\times$ELG is very noisy. This is likely due to the \civ forest sample in SB1 having little to no overlap with the ELG tracer redshifts. The \siiv region is only visible in quasar spectra with $z > 1.6$ and the ELG redshifts used in this work are all at redshifts $z < 1.6$. Though it is included in this work, we recommend that any future studies should exclude the CIV(SB1)$\times$ELG cross-correlation unless a better overlap in redshift is achieved.

\subsection{Validation Tests}
\label{subsec:validation-tests}

In this work, we follow the procedure for a well-developed method for measuring BAO with the Lyman-$\alpha$ forest. Though this method has been carefully tested and validated starting over a decade ago with SDSS, there have been few studies using this procedure with an absorber other than the Lyman-$\alpha$ forest, and even fewer studies using \civ as an absorber. In fact, the latest measurement of BAO from the \civ forest comes from \cite{blomqvist2018} which measured BAO from SDSS DR14, though they did not get a statistically significant detection. There have been no previous studies that cross-correlate ELGs with \civ flux fluctuations (absorption). 

Given the much larger dataset of DESI DR2, any validation tests would normally be run on synthetic datasets, or mocks. Unfortunately, at the time of this paper, there were no synthetic \civ forest datasets available and we were unable to validate any of our results using mock datasets. Instead, we apply validation tests to the data. 

We were able to apply a blinding scheme during all measurements and validation tests. Since this work was performed after the release of the \lyabao, we use a different blinding scheme than what was used in \lyabao. There, they applied a blinding scheme at the level of the correlations themselves. Because the pipeline no longer automatically blinded the correlations, in this work we instead apply our blinding scheme to the fits where we randomly shift the BAO parameters ($\alpha_\parallel$,~$\alpha_\perp$,~$\alpha_{\rm iso}$) by a pre-determined seeded amount. This is the same blinding scheme that was used in the DESI DR1 Lyman-$\alpha$ forest full-shape analysis \citep{cuceu2025}. Only after obtaining and understanding the results from the validation tests were the baseline results unblinded. Our goal for unblinding was that any shifts to $\alpha_{\rm iso}$ from the analysis variations would be less than $\frac{1}{3}\sigma_{\rm baseline}$ and that all fits would be valid. 

We run two types of validation tests on our analysis. The first is different analysis variations where we explore how various changes in the method affect the BAO results. These variations are discussed in section \ref{subsec:variations}. The second is data splits, where the dataset is split into two based on some criteria. In this work, we only study one data split based on the equivalent width of the \civ line and we discuss the results in section \ref{subsec:data_splits}.

\subsubsection{Analysis Variations}
\label{subsec:variations}

\begin{figure*}
    \centering
    \includegraphics[width=\linewidth]{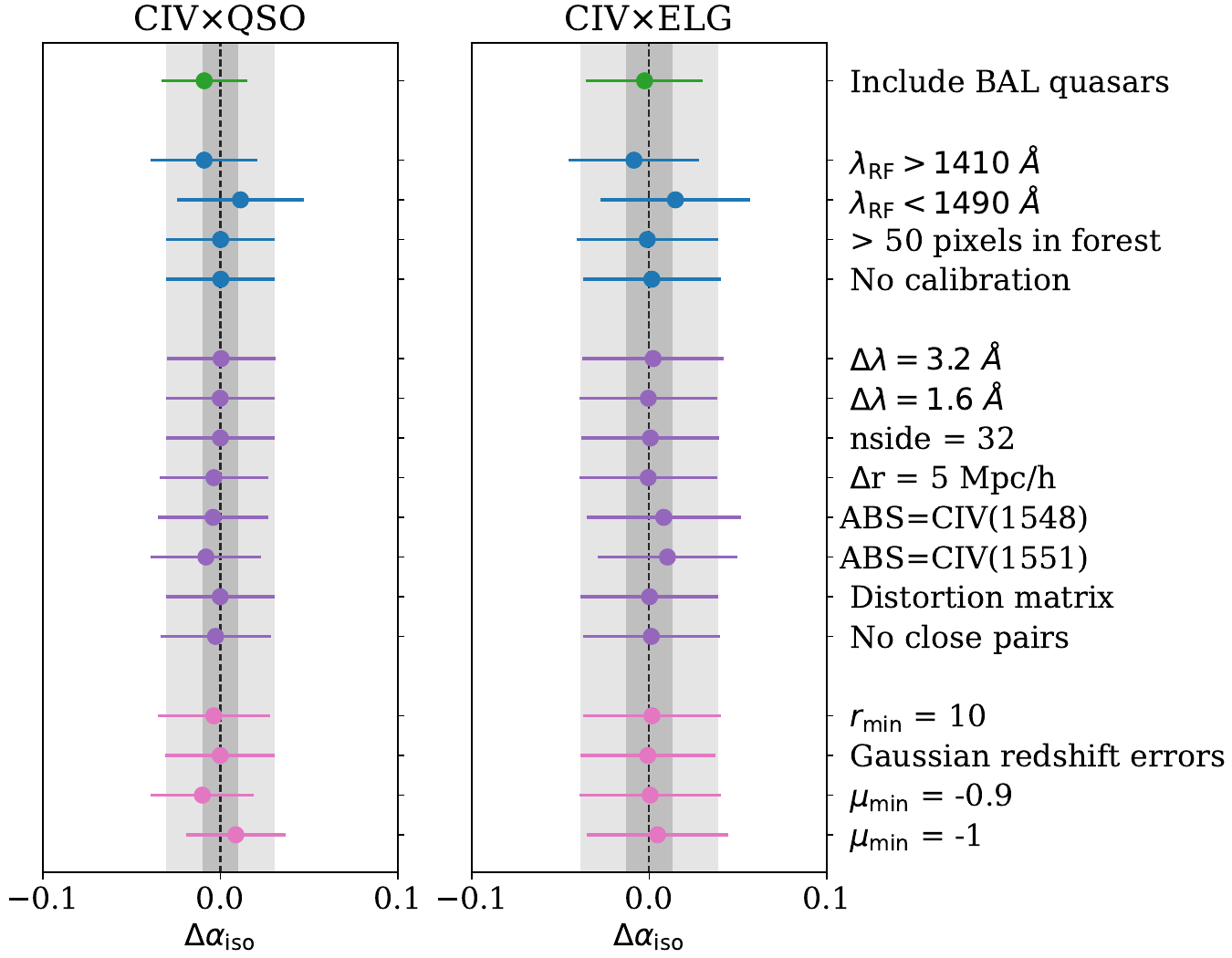}
    \caption{Shifts in the isotropic BAO parameter $\alpha_{\rm iso}$ from the different analysis variations run. Each analysis variation changes one part of the analysis. The green point corresponds to changes to the dataset; the blue points correspond to changes to the continuum fitting method; the purple points correspond to changes in the calculation of the correlations; and the pink points correspond to changes in the fit. The dark and light grey regions correspond to $\frac{1}{3}\sigma_{\rm baseline}$ and $\sigma_{\rm baseline}$, respectively. All variations for both tracers are within the $\frac{1}{3}\sigma_{\rm baseline}$ goal except for the ``\texttt{$\lambda_{\rm RF}$ < 1490 \AA}'' variation. This shift is within the expected statistical shift from using a subset of the baseline dataset.
    }
    \label{fig:variations}
\end{figure*}

We run the same analysis variations on both the quasar and ELG measurements and use the same fit parameters as in the baseline fit with small adjustments where appropriate. We present the results for the combined fits in Figure \ref{fig:variations}, which shows the one-dimensional shifts in $\alpha_{\rm iso}$ for all analysis variations. The variations are grouped based on which part of the analysis is changed. The green point is a change to the dataset:
\begin{itemize}
    \item \texttt{include BAL quasars:} we use the quasar catalog from \lyabao that includes BAL quasars. The BAL quasars are used as tracers and BAL regions are masked during the continuum fitting process following \cite{martini2025};
\end{itemize}
the blue points are changes to the continuum fitting method:
\begin{itemize}
    \item \texttt{$\lambda_{\rm RF}$ > 1410\AA:} we extend the length of the \civ region from 1420\AA ~to 1410\AA. The \siiv region remains the same as in baseline,
    \item \texttt{$\lambda_{\rm RF}$ < 1490\AA:} we shorten the length of the \civ region from 1520\AA ~to 1490\AA. The \siiv region remains the same as in baseline,
    \item \texttt{>50 pixels in forest:} we include lines-of-sight in both forest regions with more than 50 valid forest pixels. For the baseline analysis we required at least 90 pixels in SB2 and 110 pixels in SB1,
    \item \texttt{no calibration:} we do not perform the recalibration step described in \cite{cesar2023};
\end{itemize}
the purple points are changes to the correlation calculation:
\begin{itemize}
    \item \texttt{$\Delta\lambda$ = 3.2\AA:} we re-bin both forest samples by 4 pixels rather than by 3 pixels ($\Delta\lambda = 2.4$\AA),
    \item \texttt{$\Delta\lambda$ = 1.6\AA:} we re-bin both forest samples by 2 pixels rather than by 3 pixels,
    \item \texttt{nside = 32:} we measure the correlations in \texttt{HEALPix} pixels with nside=32 rather than nside=16 as in the baseline analysis,
    \item \texttt{$\Delta$r = 5 Mpc/h:} we use a 5\mpc binning of the correlations and distortion matrices rather than the 4\mpc binning in baseline,
    \item \texttt{ABS=CIV(1548):} we use the $\lambda_{\rm RF} = 1548.20$\AA~line of the doublet as the absorber instead of the average wavelength,
    \item \texttt{ABS=CIV(1551):} we use the $\lambda_{\rm RF} = 1551.78$\AA~line of the doublet as the absorber instead of the average wavelength,
    \item \texttt{distortion matrix:} we change the percent of the dataset used to calculate the distortion matrix. For the quasar measurement we use 2\% of the dataset rather than 1\%. For the ELG measurement we use 0.5\% of the dataset rather than 0.1\%,
    \item \texttt{no close pairs:} we remove quasar-pixel (or ELG-pixel) pairs that could potentially contaminate the correlations on small-scales. We use an extreme cut of 0.33 degrees and 10,000 km/s. This cut is applied to pairs in the correlations and distortion matrices;
\end{itemize}
and the pink points are changes to the fit:
\begin{itemize}
    \item \texttt{$r_{\rm min}$ = 10:} we change the minimum distance in the fit from 30\mpc to 10\mpc,
    \item \texttt{Gaussian redshift errors:} we model quasar redshift errors with a Gaussian model rather than the Lorentzian model used in the baseline analysis,
    \item \texttt{$\mu_{\rm min}$ = -0.9:} we change the angular wedge used in the fit from -0.95 to -0.9 to restrict the $\mu$ range further,
    \item \texttt{$\mu_{\rm min}$ = -1.0:} we change the angular wedge used in the fit from -0.95 to -1.0 to allow the full $\mu$ range. 
\end{itemize}
The dark and light grey regions in Figure \ref{fig:variations} correspond to $\frac{1}{3}\sigma_{\rm baseline}$ and $\sigma_{\rm baseline}$, respectively. 

For the variations where we choose one line of the doublet as the absorber (\texttt{ABS=CIV(1548)} and \texttt{ABS=CIV(1551)}), we add an additional free parameter to the fit (either $b_{\rm CIV(1551)}$ or $b_{\rm CIV(1548)}$, respectively) to model the other line of the doublet as a metal contaminant. To preserve the extracted BAO information, we also need to include the BAO model for the cross-correlation with the line of the doublet that is treated as a metal contaminant.

For both the quasar and ELG measurements, all but one of the variations are within $\frac{1}{3}\sigma_{\rm baseline}$ and the error-bars are of similar magnitude to the baseline fit. The variation that does not meet the goal is $\lambda_{\rm RF} < 1490$~\AA~and the shifts to $\alpha_{\rm iso}$ are 0.011 for quasars and 0.014 for ELGs. This variation uses a subset of the baseline analysis by restricting the length of the \civ region. We can calculate the expected statistical shift for each measurement and find that it is around 0.04 for both tracers, showing that this shift is within the expected statistical fluctuation.

\subsubsection{Data Splits}
\label{subsec:data_splits}
We study the effects on our BAO results when splitting our quasar catalog into two based on the equivalent width (EW) of the \civ doublet. A similar analysis was performed in \cite{desi-dr1-kp6}. This split is motivated by the Baldwin Effect \citep{baldwin}, which is the anti-correlation between the EW of quasar emission lines and the luminosity of the continuum. Due to this, we expect the shape of the quasar spectral energy distribution (SED) to depend on the \civ EW. 

The equivalent width information was taken from the \texttt{FastSpecFit}\footnote[3]{\texttt{FastSpecFit} is a stellar continuum and emission-line modeling code designed to model DESI data. More information can be found at \url{https://fastspecfit.readthedocs.io/}.} \citep{fastspecfit} catalogs and added to our quasar catalog. The quasar catalog is then split at the median equivalent width of 48 \AA. We again remove all BAL quasars. 

The results on $\alpha_{\rm iso}$ for the two data splits are: 
\begin{equation}
    \begin{split}
        \alpha_{\rm iso} & = 1.081 \pm 0.190 ~\text{(\civ} > 48\text{\AA)} \\
        \alpha_{\rm iso} & = 1.142 \pm 0.054 ~\text{(\civ} < 48\text{\AA)}. 
    \end{split}
\end{equation}
These results are consistent with each other and with the combined quasar result. The errorbars of the data splits did increase compared to the main result, but this is not unexpected, as this data split cuts a measurement that was already not very strong into two even weaker measurements.

\subsection{Cosmology Results}

\begin{figure*}
    \centering
    \includegraphics[width=\linewidth]{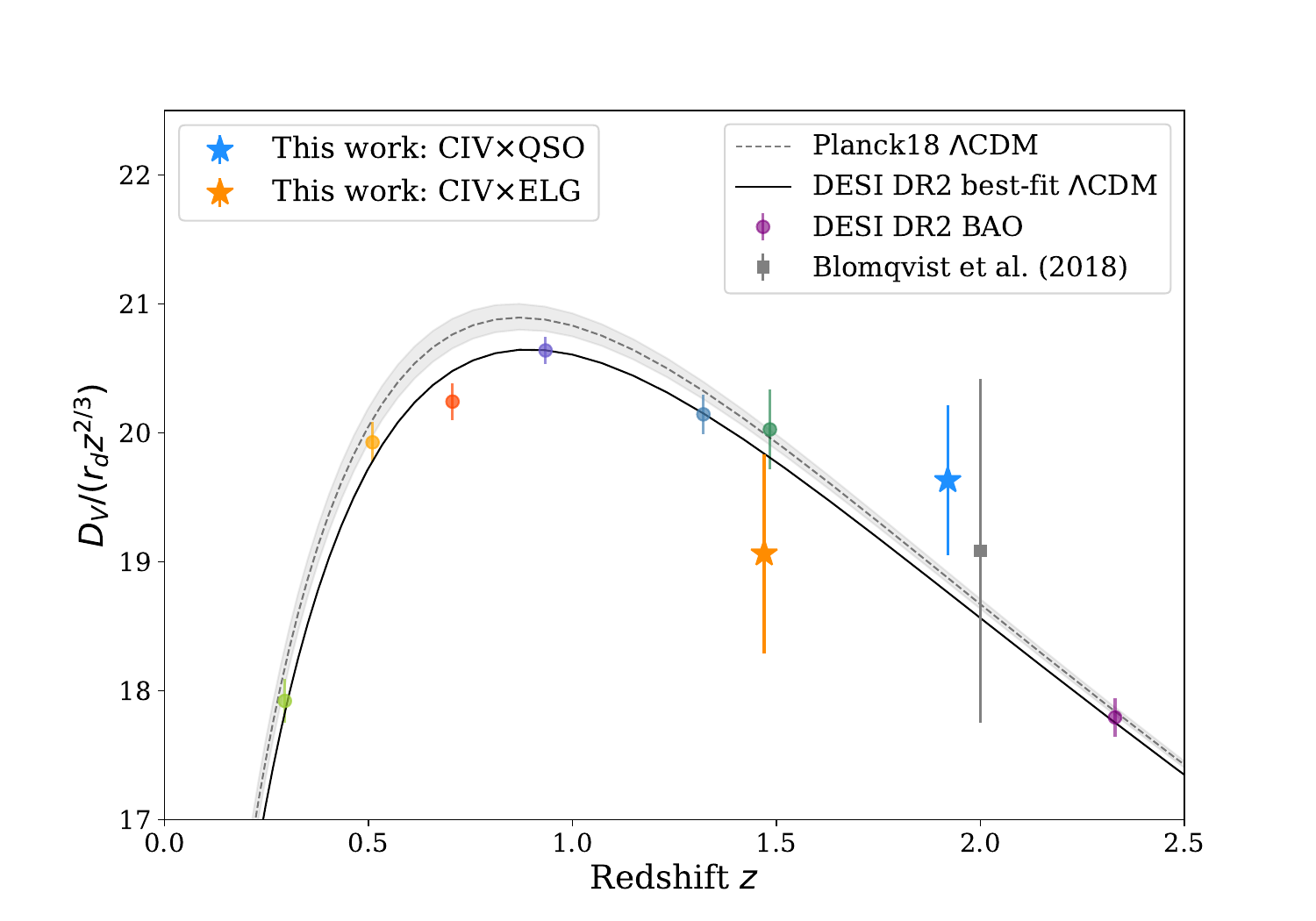}
    \caption{Measurements of the BAO distance parametrized as the ratio of the angle-averaged distance $D_{\rm V}$ and the sound horizon $r_{\rm d}$. We apply a scaling of $z^{-2/3}$ for visualization purposes. The measurement from the combined fit with quasars from this work is shown as the blue star at $z_{\rm eff} = 1.92$. We show the measurement from the combined fit with ELGs for completeness as the orange star. The measurements from the various tracers from \galbao are shown as the multi-colored points (from left to right: BGS, LRG, LRG, LRG+ELG, ELG, QSO, LY$\alpha$). The solid black line is the best-fit $\Lambda$CDM model from \galbao and the dashed grey line is the $\Lambda$CDM model from \cite{planck2018}. The grey square shows the result from \cite{blomqvist2018}.}
    \label{fig:dv}
\end{figure*}

Figure \ref{fig:dv} shows how the BAO measurements from this work compare to the results and best-fit $\Lambda$CDM model from \galbao and from \cite{planck2018}. The BAO measurements are parametrized as the ratio of the angle-averaged distance $D_{\rm V}$ (from equation \ref{eq:dv}) to the sound horizon at the baryon drag epoch, $r_{\rm d}$. For the combined fit with quasars, we find 
\begin{equation}
        D_{\rm V}/r_{\rm d}(z_{\rm eff} = 1.92) = 30.3 \pm 0.9,
\end{equation}
shown as the blue star in Figure \ref{fig:dv}, is statistically consistent with the best fit $\Lambda$CDM results from \galbao.

We present the measurement with ELGs (orange star in Figure \ref{fig:dv}) for completeness and we find 
\begin{equation}
    D_{\rm V}/r_{\rm d} (z_{\rm eff} = 1.47) = 24.6 \pm 1.0.
\end{equation}
The values shown in Figure \ref{fig:dv} are scaled by $z^{-2/3}$ for visualization purposes.

\section{Summary}
\label{sec:summary}

We present the first DESI measurement of BAO using cross-correlations of the \civ forest fluctuations with quasars and a proof of concept measurement of the cross-correlation of the \civ forest with ELGs. These measurements are based on the first three years of DESI main survey data which will be released with data release 2 (DR2). 

We find a 4.2$\sigma$ detection of BAO from the CIV$\times$QSO cross-correlation and a corresponding uncertainty on $\alpha_{\rm iso}$ of 3.1\% at an effective redshift $z_{\rm eff} = 1.92$. We find a corresponding isotropic distance measurement of $D_{\rm V}/r_{\rm d} = 30.3 \pm 0.9$. 

We measure the BAO signal at 2.5$\sigma$ significance in the CIV$\times$ELG cross-correlation. The uncertainty on $\alpha_{\rm iso}$ is 4\% at an effective redshift of $z_{\rm eff} = 1.47$. The isotropic distance measurement is $D_{\rm V}/r_{\rm d} = 24.6 \pm 1.0$.

We validated our measurements by performing several analysis variations for both tracers to test how small changes to the method affect the result on $\alpha_{\rm iso}$. We set a goal that $\Delta\alpha_{\rm iso} < \frac{1}{3}\sigma_{\rm baseline}$and all but one of the variations met this goal. This variation used a subset of the baseline dataset by shortening the length of the \civ region (changing the limit from 1520 \AA~to 1490~\AA). We confirmed that the larger shift was within the expected statistical fluctuation for this smaller dataset. We also performed one data split based on the equivalent width of the \civ line. The two splits were in agreement with each other and with the baseline quasar result. We did not perform this data split on the ELG measurement.

Future work will look to improve the uncertainty on the CIV$\times$QSO cross-correlation and increase the significance of the CIV$\times$ELG cross-correlation detection. The main way to do this is by increasing the signal in the \civ and \siiv regions of quasar spectra which, in turn, will increase the signal in the \civ forest samples. For Lyman-$\alpha$ studies, this is achieved by observing quasars ($z > 2.1$) multiple times. In DESI DR2, only 15.8\% of quasars between $1.4 < z < 2.1$ were observed more than once, shrinking to just 3.4\% for those observed more than twice. We estimate that we can reduce our error on $\alpha_{\rm iso}$ by a factor of two by observing low-redshift quasars ($1.4 < z < 2.1$) four times.

One immediate improvement on the uncertainty could be obtained by including BAL quasars. BALs are more likely to be identified in spectra with a higher signal-to-noise ratio (SNR). By excluding the BAL quasars from our \civ forest samples, we discarded spectra that had higher signal in the \civ and \siiv regions. Our validation test where we included BAL quasars confirms that we can improve the uncertainty on $\alpha_{\rm iso}$ by including these quasars. When including BALs, the error on $\alpha_{\rm iso}$ for the CIV$\times$QSO measurement is only 2.4\% and nearly 3.2\% for the ELG$\times$QSO measurement. This validation test was performed with the BAL masking strategy described in \cite{martini2025}, but future studies could decide if there are any improvements to be made. 

There are many future studies that could be conducted on how the \civ forest samples are obtained. In this work, we used \texttt{picca} to obtain the unabsorbed continuum in the \civ and \siiv regions. \texttt{picca} was designed to measure the continuum in the Lyman-$\alpha$ region, where the absorption due to Lyman-$\alpha$ is incredibly dense. Given that the \civ absorption in the \civ and \siiv regions is significantly less dense, one could look into other methods (such as spline fitting \citep{pieri2014a}) to estimate the continuum. Another continuum prediction one could use is LyCAN, a convolutional neural network that predicts the unabsorbed quasar continuum within the rest-frame wavelength range of 1040 \AA~to 1600 \AA ~\citep{turner2024}. Other studies could look at whether or not \civ absorption could be extracted from ELG spectra, and if so, whether combining these forests with the forest samples in this work could help improve the uncertainty on $\alpha_{\rm iso}$.

We leave these studies as potential future work, but reiterate that with improvements to the uncertainties, this work has the potential to provide a key BAO measurement for DESI at an effective redshift that the current tracers do not cover.

\section*{Data Availability}
{The data used in this analysis will be made public along with Data Release 2 (\url{https://data.desi.lbl.gov/doc/releases/}). The data points for reproducing figures will be made available on Zenodo upon journal publication. 
}

\section*{Acknowledgments}
{AC acknowledges support provided by NASA through the NASA Hubble Fellowship grant HST-HF2-51526.001-A awarded by the Space Telescope Science Institute, which is operated by the Association of Universities for Research in Astronomy, Incorporated, under NASA contract NAS5-26555.

This material is based upon work supported by the U.S. Department of Energy (DOE), Office of Science, Office of High-Energy Physics, under Contract No. DE–AC02–05CH11231, and by the National Energy Research Scientific Computing Center, a DOE Office of Science User Facility under the same contract. Additional support for DESI was provided by the U.S. National Science Foundation (NSF), Division of Astronomical Sciences under Contract No. AST-0950945 to the NSF’s National Optical-Infrared Astronomy Research Laboratory; the Science and Technology Facilities Council of the United Kingdom; the Gordon and Betty Moore Foundation; the Heising-Simons Foundation; the French Alternative Energies and Atomic Energy Commission (CEA); the National Council of Humanities, Science and Technology of Mexico (CONAHCYT); the Ministry of Science, Innovation and Universities of Spain (MICIU/AEI/10.13039/501100011033), and by the DESI Member Institutions: \url{https://www.desi.lbl.gov/collaborating-institutions}. Any opinions, findings, and conclusions or recommendations expressed in this material are those of the author(s) and do not necessarily reflect the views of the U. S. National Science Foundation, the U. S. Department of Energy, or any of the listed funding agencies.

The authors are honored to be permitted to conduct scientific research on I'oligam Du'ag (Kitt Peak), a mountain with particular significance to the Tohono O’odham Nation.
}

\bibliographystyle{aasjournal}
\bibliography{references}

\appendix

\section{Baseline Fit Results}
\label{sec:app_b}

\begin{figure}[H]
    \centering
    \includegraphics[width=0.95\linewidth]{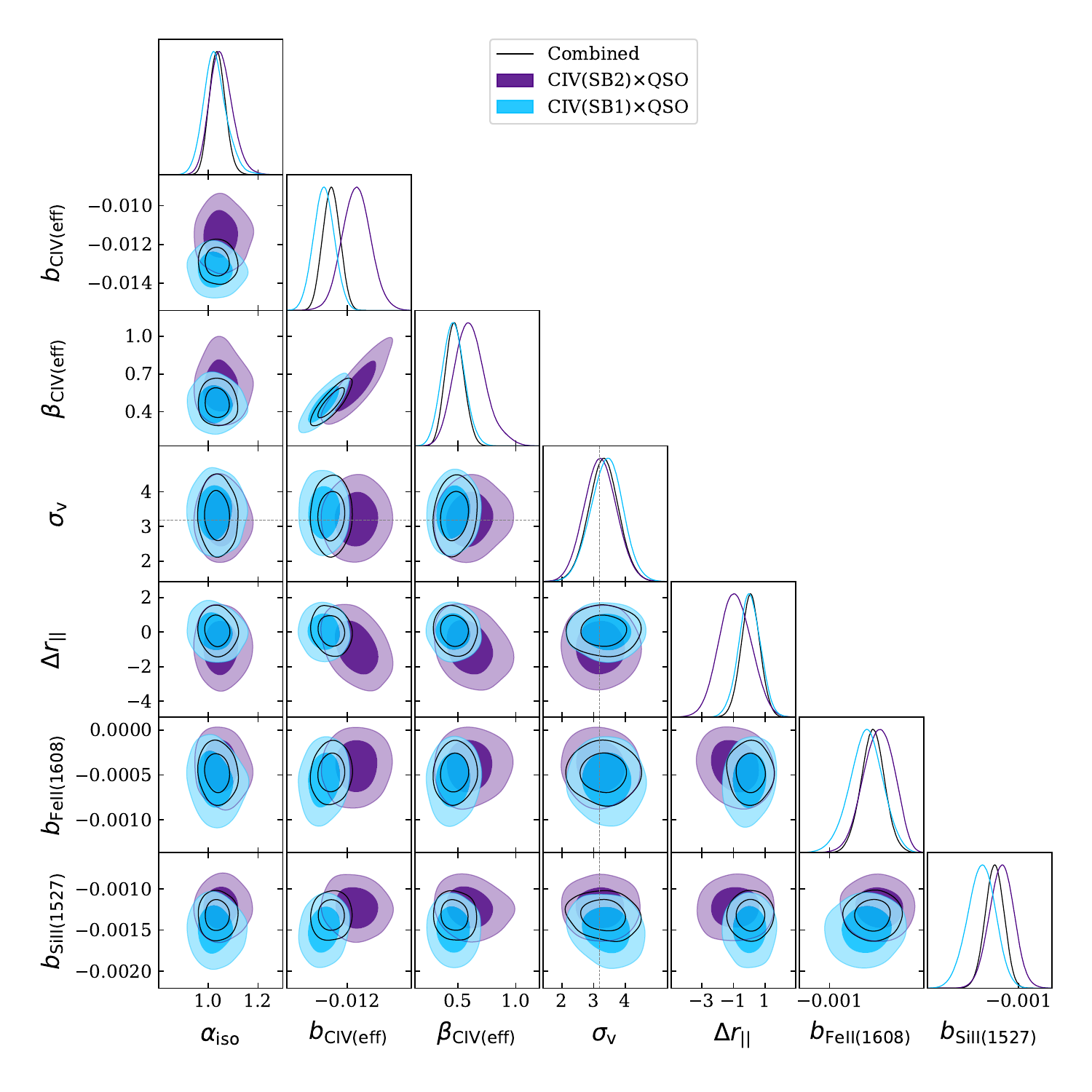}
    \caption{Results from running the sampler on the CIV(SB2)$\times$QSO (purple), CIV(SB1)$\times$QSO (blue), and combined (black) fits. }
    \label{fig:triangle-baseline-qso}
\end{figure}

\begin{figure}[H]
    \centering
    \includegraphics[width=0.95\linewidth]{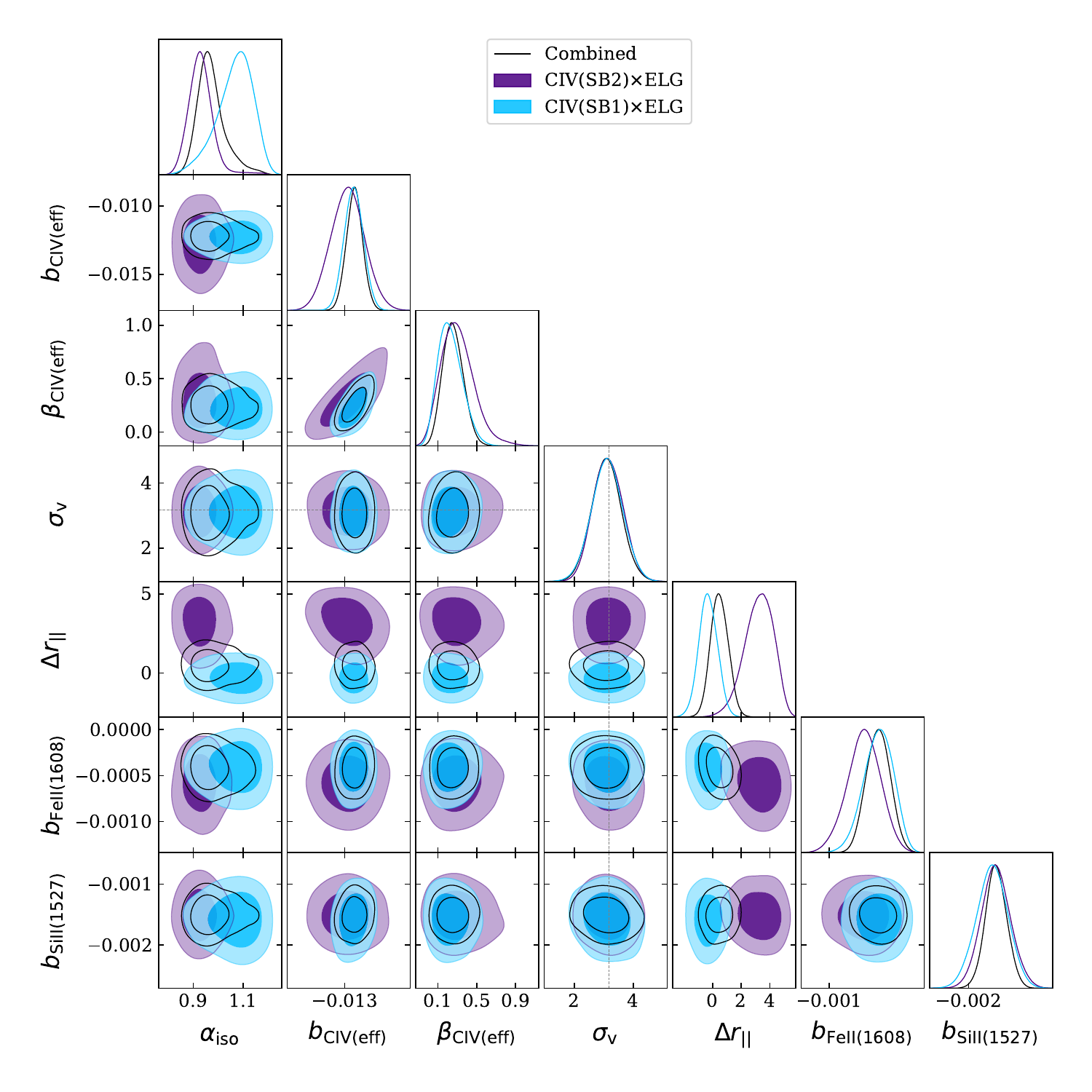}
    \caption{Results from running the sampler on the CIV(SB2)$\times$ELG (purple), CIV(SB1)$\times$ELG (blue), and combined (black) fits.}
    \label{fig:triangle-baseline-elg}
\end{figure}

We show contour plots from our baseline fit from running the Polychord sampler on our correlations in Figures \ref{fig:triangle-baseline-qso} and \ref{fig:triangle-baseline-elg}. Figure \ref{fig:triangle-baseline-qso} shows the results from the combined fit with quasars (black line contours), as well as fits to the individual side bands (purple and blue contours). Figure \ref{fig:triangle-baseline-elg} shows the analogous results for ELGs. The contours from the three fits for each tracer are in good agreement.

\end{document}